\documentclass{llncs}

\usepackage{amsmath,amssymb,graphics}
\usepackage{gdefs}

\usepackage{graphicx}
\usepackage{paralist}

\usepackage{epsfig,amsmath,gdefs,graph,alltt,ifthen,times,paralist}
\usepackage{graphics}
\usepackage{graphicx}

\usepackage{setspace}
\usepackage{changebar}
\usepackage{rotating}
\usepackage{latexsym}

\newcommand{\defref}[1]{Def.~\ref{def:#1}}
\newcommand{\lemref}[1]{Lemma~\ref{lem:#1}}
\newcommand{\theref}[1]{Theorem~\ref{the:#1}}
\newcommand{\secref}[1]{Section~\ref{Sec:#1}}
\newcommand{\appref}[1]{Appendix~\ref{app:#1}}
\newcommand{\figref}[1]{Fig.~\ref{Fig:#1}}
\newcommand{\tableref}[1]{Table~\ref{Ta:#1}}
\newcommand{\equref}[1]{(\ref{eq:#1})}

\newcommand{\pred}[3]{{#1}{\RBox{#2}}{#3}}

\newcommand{\negFormula}[3]{{#1}{\RBox{#2}}{#3}}
\newcommand{\posFormula}[3]{{#1}{\RDiamond{#2}}{#3}}

\newcommand{\LRP}{\mbox{\textit{LRP}}}
\newcommand{\LRPtwo}{\LRP_2}
\newcommand{\NLBPtwo}{C\LRPtwo}

\renewcommand{\phi}{\varphi}
\newcommand{\gaifman}[1]{\mathcal{G}({#1})}

\newcommand{\Hide}[1]{}

\newcommand{\Ayah}[1]{\mathcal{A}_{#1}}
\newcommand{\Tr}{Tr}

\newcommand{\Stam}[1]{}

\newcommand{\voc}{\tau}

\newcommand{\ktree}{T^{k}}

\newcommand{\blank}{\flat}

\renewcommand{\implies}{\Rightarrow}

\newboolean{TR}
\setboolean{TR}{false}
\ifthenelse{\boolean{TR}}{
\newcommand{\TrSelect}[2]{#1}
\newcommand{\TrOnly}[1]{#1}
\newcommand{\SubOnly}[1]{}
\newcommand{\TrOnlyInFootnote}[1]{#1}
\newcommand{\TrOnlyInTable}[1]{#1}}
{
\newcommand{\TrSelect}[2]{#2}
\newcommand{\TrOnly}[1]{}
\newcommand{\SubOnly}[1]{#1}
\newcommand{\TrOnlyInFootnote}[1]{}
\newcommand{\TrOnlyInTable}[1]{}}

\input{xy}
\xyoption{all}

\begin{document}

\title{A Logic of Reachable Patterns \\in Linked Data-Structures}

\author{Greta Yorsh\inst{1}\thanks{This research was supported by THE ISRAEL SCIENCE FOUNDATION (grant No 304/03).}
\and Alexander Rabinovich\inst{1} \and Mooly Sagiv\inst{1}
\and \\ Antoine Meyer\inst{2} \and Ahmed Bouajjani\inst{2}}

\institute{
Tel Aviv Univ., Israel. $\{$gretay,rabinoa,msagiv$\}$@post.tau.ac.il
\and
Liafa, Univ. of Paris 7, France.
$\{$ameyer,abou$\}$@liafa.jussieu.fr
}

\maketitle

\nochangebars{
\begin{abstract}
We define a new decidable logic for expressing and checking invariants of programs
that manipulate dynamically-allocated objects via pointers and destructive pointer updates.
The main feature of this logic is the ability to limit the neighborhood of a node
that is reachable via a regular expression from a designated node.
The logic is closed under boolean operations (entailment, negation) and has a finite model property.
The key technical result is the proof of decidability.

We show how to express precondition, postconditions, and loop invariants for some interesting programs.
It is also possible to express properties such as disjointness of data-structures, and low-level heap mutations.
Moreover, our logic can express properties of arbitrary data-structures and of an arbitrary number of pointer fields.
The latter provides a way to naturally specify postconditions that relate the fields on entry to
a procedure to the fields on exit.
Therefore, it is possible to use the logic to automatically prove partial correctness of programs
performing low-level heap mutations.
\end{abstract}

\section{\label{Se:Intro}Introduction}

The automatic verification of programs with dynamic memory allocation and
pointer manipulation is a challenging problem.
In fact, due to dynamic memory allocation and destructive updates of pointer-valued fields,
the program memory can be of arbitrary size and structure.
This requires the ability to
reason about a potentially infinite number of memory (graph) structures,
even for programming languages that have good capabilities for data abstraction.
Usually abstract-datatype operations are implemented using loops,
procedure calls, and sequences of low-level pointer manipulations;
consequently, it is hard to prove that a data-structure invariant is
reestablished once a sequence of operations is finished \cite{kn:Hoare75}.

To tackle the verification problem of such complex programs,
several approaches emerged in the last few years
with different expressive powers and levels of automation,
including works based on abstract interpretation \cite{SAS:LS00,TOPLAS:SRW02,CAV:RSW04},
logic-based reasoning \cite{POPL:OI01,Rey02}, and automata-based techniques \cite{KS93,PLDI:MS01,BHMV05}.
An important issue is the definition of a formalism that (1) allows us to express
relevant properties (invariants) of various kinds of linked data-structures, and (2) has the
closure and decidability features needed for automated verification.
The aim of this paper is to study such a formalism based on logics over arbitrary graph structures,
and to find a balance between expressiveness, decidability and complexity.

Reachability is a crucial notion for reasoning about linked data-structures.
For instance, to establish that a memory configuration contains no
garbage elements, we must show that every element is reachable from
some program variable. Other examples of properties that involve reachability are
(1)~the acyclicity of data-structure fragments, i.e.,
every element reachable from node $u$ cannot reach $u$,
(2)~the property that a data-structure traversal terminates, e.g., there is
a path from a node to a sink-node of the data-structure,
(3)~the property that, for programs with procedure calls when references are passed as arguments,
elements that are \emph{not} reachable from a formal parameter are not modified.

A natural formalism to specify properties involving reachability is the first-order logic
over graph structures with transitive closure.
Unfortunately, even simple decidable fragments of first-order logic
become undecidable when transitive closure is added \cite{GOR99,CSL:eadtc}.

In this paper, we propose a logic that can be seen as a fragment
of the first-order logic with transitive closure.
Our logic is (1)~simple and natural to use,
(2)~expressive enough to cover important properties of a wide class of arbitrary linked
data-structures, and
(3)~allows for algorithmic modular verification using programmer's specified
loop-invariants and procedure's specifications.

Alternatively, our logic can be seen as a propositional logic with atomic proposition modelling
reachability between heap objects pointed-to by program variables and other heap objects with
certain properties.
The properties are specified using patterns that limit the neighborhood of an object.
For example, in a doubly linked list, a pattern says that if an object $v$ has an
an emanating \texttt{forward} pointer that leads to an object $w$,
then $w$ has a \texttt{backward} pointer into $v$.

The contributions of this paper can be summarized as follows:
\begin{itemize}
\item We define the {\em Logic of Reachable Patterns} ($\LRP$)
where reachability constraints such as those mentioned
above can be used.
Patterns in such constraints are defined by quantifier-free first-order
formulas over graph structures and sets of access paths are defined by regular expressions.
\item We show that $\LRP$
has a finite-model property, i.e., every satisfiable formula has a finite model.
Therefore, invalid formulas are always falsified by a finite store.
\item We prove that the logic $\LRP$ is, unfortunately, undecidable.
\item We define a suitable restriction on the patterns
leading to a fragment of $\LRP$ called $\LRPtwo$.
\item We prove that the satisfiability (and validity) problem is decidable.
The fragment $\LRPtwo$ is the main technical result of the paper and the decidability proof is
non-trivial.
The main idea is to show that every satisfiable $\LRPtwo$ formula is also satisfied by a tree-like graph.
Thus, even though $\LRPtwo$ expresses properties of arbitrary data-structures, because the logic
is limited enough, a formula that is satisfied on an arbitrary graph is also satisfied on a tree-like graph.
Therefore, it is possible to answer satisfiability (and validity) queries for $\LRPtwo$ using
a decision procedure for monadic second-order logic (MSO) on trees.
\item We show that despite the restriction on patterns we introduce,
the logic $\LRPtwo$ is still expressive enough for use in program verification:
various important data-structures,
and loop invariants concerning their manipulation, are in fact definable in $\LRPtwo$.
\end{itemize}

The new logic $\LRPtwo$ forms a basis of the verification framework
for programs with pointer manipulation~\cite{tr05:VC},
which has important advantages w.r.t. existing ones.
For instance, in contrast to decidable logics that restrict the graphs of interest
(such as monadic second-order logic on trees), our logic allows arbitrary graphs
with an arbitrary number of fields.
We show that this is very useful even for verifying programs that manipulate singly-linked lists
in order to express postcondition and loop invariants that relate the input and the output
state.
Moreover, our logic strictly generalizes the decidable logic in \cite{ESOP:BRS99}, which inspired our work.
Therefore, it can be shown that certain heap abstractions including~\cite{kn:Hendren,kn:SRW98}
can be expressed using $\LRPtwo$ formulas.

The rest of the paper is organized as follows:
\secref{Logic} defines the syntax and the semantics of $\LRP$,
and shows that it has a  finite model property, and that $\LRP$ is undecidable;
\secref{express} defines the fragment $\LRPtwo$, and
demonstrates the expressiveness of $\LRPtwo$ on
several examples;
\secref{decidable} describes the main ideas of the decidability proof for $\LRPtwo$;
\secref{LimitExtend} discusses the limitations and the extensions of the new logics;
finally, \secref{related} discusses the related work.
\TrSelect{Proofs of technical lemmas are given in the appendix.}
{The full version of the paper~\cite{tr05:LRP} contains
the formal definition of the semantics of $\LRP$ and proofs.}

\newcommand{\RBox}[1]{[#1]}
\newcommand{\RDiamond}[1]{\B{#1}}
\newcommand{\Edge}[1]{\buildrel #1 \over \to}
\newcommand{\BEdge}[1]{\buildrel #1 \over \leftarrow}
\newcommand{\ci}{{\cal c}}
\newcommand{\ui}{{\cal u}}
\newcommand{\All}{\textbf{A}}
\newcommand{\Exists}{\textbf{E}}

\section{The $\LRP$ Logic}\label{Sec:Logic}

In this section, we define the syntax and the semantics of our logic.
For simplicity, we explain the material in terms of expressing properties of heaps.
However, our logic can actually model properties of arbitrary directed graphs.
Still, the logic is powerful enough to express the property that a graph denotes a heap.

\subsection{Syntax of $\LRP$}

$\LRP$ is a propositional logic over reachability constraints.
That is, an $\LRP$ formula is a boolean combination of closed formulas in first-order logic with
transitive closure that satisfy certain syntactic restrictions.

Let $\voc = \B{C, U, F}$  denote a vocabulary, where
(i)~$C$ is a finite set of constant symbols usually denoting designated objects in the heap,
pointed to by program variables;
(ii)~$U$ is a set of unary relation symbols denoting properties, e.g., color of a node in a Red-Black tree;
(ii)~$F$ is a finite set of binary relation symbols (edges) usually denoting pointer fields.\footnote{We
can also allow auxiliary constants and fields including abstract fields~\cite{JML}.}

A \textbf{term\/} $t$ is either a variable or a constant $c \in C$.
An \textbf{atomic formula} is an equality $t = t'$, a unary relation $u(t)$,
or an edge formula $t \Edge{f} t'$, where $f \in F$, and $t, t'$ are terms.
A \textbf{quantifier-free formula} $\psi(v_0, \ldots, v_n)$ over $\voc$ and
variables $v_0, \ldots, v_n$
is an arbitrary boolean combination of atomic formulas.
Let $FV(\psi)$ denote the free variables of the formula $\psi$.

\begin{definition}\label{def:gaifman}
Let $\psi$ be a conjunction of edge formulas of the form $v_i \Edge{f} v_j$,
where $f \in F$ and $0 \leq i,j \leq n$.
The \textbf{Gaifman graph} of $\psi$, denoted by
$B_{\psi}$, is an undirected graph with a vertex for each free variable of $\psi$.
There is an arc between the vertices corresponding to $v_i$ and $v_j$ in $B_{\psi}$ if and only if
$(v_i \Edge{f} v_j)$ appears in $\psi$, for some $f \in F$.
The \textbf{distance} between logical variables $v_i$ and $v_j$ in the formula $\psi$
is the minimal edge distance between the corresponding
vertices $v_i$ and $v_j$ in $B_{\psi}$.
\end{definition}
For example, for the formula $\psi = (v_0 \Edge{f} v_1) \land (v_0 \Edge{f} v_2)$
the distance between $v_1$ and $v_2$ in $\psi$ is $2$,
and its underlying graph
$B_{\psi}$ looks like this:  $v_1$ --- $v_0$ --- $v_2$.

\begin{definition}\label{De:Syntax}\begin{Name}Syntax of $\LRP$\end{Name}
A \textbf{neighborhood formula} $N(v_0, \ldots, v_n)$
is a conjunction of edge formulas of the form $v_i \Edge{f} v_j$,
where $f \in F$ and $0 \leq i,j \leq n$.

A \textbf{routing expression} is an extended regular expression, defined as follows:
\[
\begin{array}{lcllr}
  R  &  ::= & \emptyset                     && \mbox{empty set}\\
     &  | &  \epsilon                     && \mbox{empty path}\\
     & | &     \Edge{f} & f \in F  & \mbox{forward along edge}\\
     & | &     \BEdge{f} & f \in F  & \mbox{backward along edge}\\
     & | &     u & u \in U  & \mbox{test if u holds}\\
     & | &     \neg u & u \in U  & \mbox{test if u does not hold}\\
     & | &     c & c \in C  & \mbox{test if c holds}\\
     & | &     \neg c & c \in C  & \mbox{test if c does not hold}\\
     &  | & R_1 . R_2 && \mbox{concatenation}\\
     & | & R_1 | R_2 && \mbox{union}\\
     & | & R^* && \mbox{Kleene star}
\end{array}
\]
A routing expression can require that a path traverse some edges backwards.
A routing expression has the ability to test
presence and absence of certain unary relations and constants along the path.

A \textbf{reachability constraint} is a closed formula of the form:
\[
\forall v_0, \ldots, v_n . R(c, v_0) \implies ( N(v_0, \ldots, v_n) \implies \psi(v_0, \ldots, v_n) )
\]
where $c \in C$ is a constant, $R$ is a routing expression, $N$ is a neighborhood formula,
and $\psi$ is an arbitrary quantifier-free formula, such that
$FV(N) \subseteq \{ v_0, \ldots, v_n \}$
and $FV(\psi) \subseteq FV(N)\cup\{ v_0 \}$.
In particular, if the neighborhood formula $N$ is $true$ (the empty conjunction),
then $\psi$ is a formula with a single free variable $v_0$.

An \textbf{$\LRP$ formula} is a boolean combination of reachability constraints.
\end{definition}
The subformula $N(v_0, \ldots, v_n) \implies \psi(v_0, \ldots, v_n)$ defines a \textbf{pattern},
denoted by $p(v_0)$.
Here, the designated variable $v_0$ denotes a ``central'' node of the ``neighborhood''
reachable from $c$ by following an $R$-path.
Intuitively, neighborhood formula $N$ binds the variables $v_0, \ldots, v_n$ to nodes that form
a subgraph, and $\psi$ defines more constraints on those nodes.~\footnote{
In all our examples, a neighborhood formula $N$ used in a pattern is such
that $B_N$ (the Gaifman graph of $N$) is connected.}

We use \textbf{let} expressions to specify the scope in which the pattern is declared:
\[
  \textbf{let}~p_1(v_0) \eqdef N_1(v_0, v_1, \ldots, v_n)\implies\psi_1(v_0, \ldots, v_n) ~\textbf{in}~\phi
\]
This allows us to write more concise formulas via sharing of patterns.

\paragraph{Shorthands}
We use $\pred{c}{R}{p}$ to denote a reachability constraint.
Intuitively, the reachability constraint requires that every node that is reachable
from $c$ by following an $R$-path satisfy the pattern $p$.

We use $\pred{c_1}{R}{\neg c_2}$ to denote
$\textbf{let}~p(v_0) \eqdef (true \implies \neg (v_0 = c_2))~\textbf{in}~\pred{c_1}{R}{p}$.
In this simple case, the neighborhood is only the node assigned to $v_0$.
Intuitively, $\pred{c_1}{R}{\neg c_2}$ means that the node labelled by constant $c_2$ is not reachable along
an $R$-path from the node labelled by $c_1$.
We use $\posFormula{c_1}{R}{c_2}$ as a shorthand for $\neg ( \pred{c_1}{R}{\neg c_2})$.
Intuitively, $\posFormula{c_1}{R}{c_2}$ means that \emph{there exists} an $R$-path from $c_1$ to $c_2$.
We use $c_1 = c_2$ to denote $\posFormula{c_1}{\epsilon}{c_2}$,
and $c_1 \neq c_2$ to denote $\neg (c_1 = c_2)$.
We use $\negFormula{c}{R}{(p_1 \land p_2)}$ to denote
$(\negFormula{c}{R}{p_1}) \land (\negFormula{c}{R}{p_2})$, when $p_1$ and $p_2$ agree
on the central node variable.
When two patterns are often used together,
we introduce a name for their conjunction (instead of naming each one separately):
$\textbf{let}~p(v_0) \eqdef (N_1\implies\psi_1) \land
 (N_2\implies\psi_2)~\textbf{in}~\phi$.

In routing expressions, we use $\Sigma$ to denote $(\Edge{f_1} | \Edge{f_2} | \ldots | \Edge{f_m})$,
the union of all the fields in $F$.
For example, $\pred{c_1}{\Sigma^*}{\neg c_2}$ means that $c_2$ is not reachable from $c_1$ by any path.
Finally, we sometimes omit the concatenation operator ``$.$'' in routing expressions.

\paragraph{Semantics}
An interpretation for an $\LRP$ formula over $\voc = \B{C, U, F}$
is a labelled directed graph $G = \B{V^G, E^G, C^G, U^G}$
where:
(i)~$V^G$ is a set of nodes modelling the heap objects, 
(ii)~$E^G \colon F \to \mathcal{P}(V^G \times V^G)$ are labelled edges,
(iii)~$C^G \colon C \to V^G$
provides interpretation of constants as unique labels on the nodes of the graph,
and
(iv)~$U^G \colon U \to \mathcal{P}(V^G)$ maps unary relation symbols to
the set of nodes in which they hold.

We say that node $v \in G$ is labelled with $\sigma$
if $\sigma \in C$ and $v = C^G(\sigma)$ or $\sigma \in U$ and $v \in U^G(\sigma)$.
In the rest of the paper, \emph{graph} denotes a directed labelled graph, in which
nodes are labelled by constant and unary relation symbols, and edges are labelled by
binary relation symbols, as defined above.

We define a satisfaction relation $\models$ between a graph $G$
and $\LRP$ formula ($G \models \phi$) similarly to the usual semantics
the first-order logic with transitive closure over graphs
\TrSelect{
(see \defref{FormulaMeaning} or translation from $\LRP$ to MSO in \appref{LRP2MSO})}
{(see~\cite{tr05:LRP})}.

\subsection{Properties of $\LRP$}

$\LRP$ with arbitrary patterns has a finite model property.
If formula $\phi \in \LRP$ has an infinite model, each reachability constraint in $\phi$
that is satisfied by this model has a finite witness.

\begin{theorem}\label{the:Finite}
\begin{Name}Finite Model Property\end{Name}
Every satisfiable $\LRP$  formula is satisfiable by a finite graph.
\begin{Sketch}
We show that $\LRP$ can be translated into a fragment of an infinitary
logic that has a finite model property.
Observe that $\pred{c}{R}{p}$ is equivalent to an infinite conjunction of universal
first-order sentences.
Therefore, if $G$ is a model of $\pred{c}{R}{p}$ then every substructure of $G$ is also its model.
Dually,  $\neg \pred{c}{R}{p}$ is equivalent to an infinite disjunction of existential first-order sentences.
Therefore, if $G$ is a model of $\neg \pred{c}{R}{p}$,
then $G$ has a finite substructure $G'$ such that every
substructure of $G$ that contains $G'$ is a model of  $\neg \pred{c}{R}{p}$.
It follows that every satisfiable boolean combination of formulas of the form  $\pred{c}{R}{p}$  has a finite model.
Thus, $\LRP$ has a finite model property.
\end{Sketch}
\end{theorem}

The logic $LRP$ is undecidable.
The proof uses a reduction from the halting problem of a Turing machine.
\begin{theorem}\label{the:Undecidable}
\begin{Name}Undecidability\end{Name}
The satisfiability problem of  $\LRP$ formulas is undecidable.
\begin{Sketch}
Given a Turing machine $M$, we construct a formula $\varphi_{M}$
such that $\varphi_{M}$ is satisfiable if and only if the execution of $M$ eventually halts.

The idea is that each node in the graph that satisfies $\varphi_{M}$ describes a cell of a tape in
some configuration, with unary relation symbols encoding the symbol in each cell,
the location of the head and the current state.
The $n$-edges describe the sequence of cells in a configuration and a sequence of configurations.
The $b$-edges describe how the cell is changed from one configuration to the next.
The constant $c_1$ marks the node that describes the first cell of the tape in the first configuration,
the constant $c_2$ marks the node that describes the first cell in the second configuration,
and the constant $c_3$ marks the node that describes the last cell in the last configuration (see sketch in \figref{model}).

\begin{figure}
\begin{center}
\includegraphics[width=1.4cm,angle=-90]{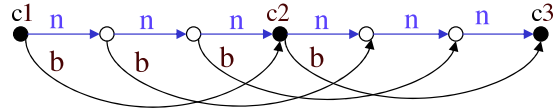}
\end{center}
\caption{\label{Fig:model} sketch of a model.}
\end{figure}

The most interesting part of the formula $\varphi_{M}$
ensures that all graphs that satisfy $\varphi_{M}$ have a grid-like form.
It states that for every node $v$ that is $n$-reachable from $c_1$, if there is a $b$-edge from $v$
to $u$, then there is a $b$-edge from the $n$-successor of $v$ to the
$n$-successor of $u$:
\begin{equation}\label{eq:undec}
\begin{array}{ll}
\textbf{let } & p(v) \eqdef (v \Edge{b} u) \land (v \Edge{n} v_1) \land (u \Edge{n} u_1)
\implies (v_1 \Edge{b} u_1)
\textbf{ in }~\negFormula{c_1}{(\Edge{n})^*}{p}
\end{array}
\end{equation}
\end{Sketch}
\end{theorem}

\begin{Remark}
The reduction uses only two binary relation symbols and a fixed number of unary relation symbols.
It can be modified to show that the logic
with three binary relation symbols (and no unary relations) is undecidable.
\end{Remark}

\section{The $\LRPtwo$ Fragment and its Usefulness}\label{Sec:express}

In this section we define the $\LRPtwo$ fragment of $\LRP$,
by syntactically restricting the patterns.
The main idea is to limit the distance between the nodes in the pattern in certain situations.

\begin{definition}\label{def:LBStwoPatterns}
A formula is in $\LRPtwo$ if in every reachability constraint $\pred{c}{R}{p}$,
with a pattern $p(v_0) \eqdef N(v_0, \ldots, v_n) \implies \psi(v_0, \ldots, v_n)$,
$\psi$ has one of the following forms:
\begin{itemize}
\item
\textbf{(equality pattern)} $\psi$ is a an equality between variables $v_i = v_j$, where $0 \leq  i,j \leq n$,
and the distance between $v_i$ and $v_j$ in $N$ is at most $2$ (distance is defined in \defref{gaifman}),
\item
\textbf{(edge pattern)} $\psi$ is of the form $v_i \Edge{f} v_j$ where $f \in F$ and $0 \leq  i,j \leq n$,
and the distance between $v_i$ and $v_j$ in $N$ is at most $1$.
\item
\textbf{(negative pattern)} atomic formulas appear only negatively in $\psi$.
\end{itemize}
\end{definition}

\begin{Remark}
Note that formula \equref{undec}, which is used in the proof of undecidability in \theref{Undecidable},
is not in $\LRPtwo$,
because $p$ is an edge pattern with distance $3$ between $v_1$ and $u_1$,
while $\LRPtwo$ allows edge patterns with distance at most $1$.
\end{Remark}

\subsection{Describing Linked Data-Structures}

In this section, we show that $\LRPtwo$ can express properties of data-structures.
\tableref{Pattern} lists some useful patterns and their meanings.
For example, the first pattern $det_f$ means that there is at most one outgoing $f$-edge from
a node.
Another important pattern $uns_f$ means that a node has at most one incoming $f$-edge.
We use the subscript $f$ to emphasize that this definition is parametric in $f$.
\begin{table*}
\begin{center}
\begin{tabular}{|l|l|l|}
\hline
\textbf{Pattern Name} & \textbf{Pattern Definition} & \textbf{Meaning}  \\
\hline
$det_f(v_0)$ & $(v_0 \Edge{f}  v_1) \land (v_0 \Edge{f} v_2) \implies (v_1 = v_2)$ &
$f$-edge from $v_0$ is deterministic\\
\hline
$uns_f(v_0)$ & $(v_1 \Edge{f} v_0) \land (v_2 \Edge{f} v_0) \implies (v_1 = v_2)$&
$v_0$ is not heap-shared by $f$-edges\\
\hline
$uns_{f,g}(v_0)$ &
$(v_1 \Edge{f} v_0) \land (v_2 \Edge{g} v_0) \implies false$
&
$v_0$ is not heap-shared by $f$-edge and $g$-edge\\
\hline
$inv_{f,b}(v_0)$ &
$\begin{array}{ll}
&(v_0 \Edge{f}  v_1 \implies v_1 \Edge{b}  v_0)\\
\land & (v_0 \Edge{b}  v_1 \implies v_1 \Edge{f} v_0)
\end{array}$
&
\begin{minipage}{2in} edges $f$ and $b$ form a doubly-linked list
between $v_0$ and $v_1$
\end{minipage}\\
\hline
$same_{f,g}(v_0)$ &
$\begin{array}{ll}
&(v_0 \Edge{f}  v_1 \implies v_0 \Edge{g}  v_1)\\
\land & (v_0 \Edge{g}  v_1 \implies v_0 \Edge{f} v_1)
\end{array}$
&
\begin{minipage}{2in}
edges $f$ and $g$ emanating from $v_0$ are parallel
\end{minipage}\\
\hline
\end{tabular}
\end{center}
\caption{\label{Ta:Pattern}%
Useful pattern definitions ($f, b, g \in F$ are edge labels).}
\end{table*}

\paragraph{Well-formed heaps}
We assume that $C$ (the set of constant symbols) contains a constant
for each pointer variable in the program (denoted by $x$, $y$ in our examples).
Also, $C$ contains a designated constant $null$ that represents \texttt{NULL} values.
Throughout the rest of the paper we assume that all the graphs denote well-formed heaps,
i.e., the fields of all objects reachable from constants are deterministic, and dereferencing NULL yields $null$.
In $\LRPtwo$ this is expressed by the formula:
\begin{equation}
(\Land_{c \in C} \Land_{f \in F} \pred{c}{\Sigma^*}{det_f})
\land (\Land_{f \in F} \posFormula{null}{\Edge{f}}{null})
\label{eq:WF}
\end{equation}

\begin{table*}
\begin{center}
\begin{tabular}{|l|l|}
\hline
\textbf{Name} & \textbf{Formula} \\
\hline
$reach_{x,f,y}$ & $\posFormula{x}{(\Edge{f})^*}{y}$ \\&
the heap object pointed-to by $y$ is reachable from the heap object pointed-to by $x$.\\
\hline
$cyclic_{x,f}$ & $\posFormula{x}{(\Edge{f})^+}{x}$ \\& cyclicity:
the heap object pointed-to by $x$ is located on a cycle.\\
\hline
$unshared_{x,f}$ &
$\negFormula{x}{(\Edge{f})^*}{uns_f}$ \\&
every heap object reachable from $x$ by an $f$-path has at most one incoming $f$-edge.\\
\hline
$disjoint_{x, f, y, g}$ & $\negFormula{x}{ (\Edge{f})^* (\BEdge{g})^*}{\neg y}$ \\&
disjointness: there is no heap object that is reachable from $x$ by an $f$-path\\&
and also reachable from $y$ by a $g$-path.\\
\hline
$same_{x,f,g}$ & $\negFormula{x}{(\Edge{f}|\Edge{g})*}{same_{f, g}}$\\&
the $f$-path and the $g$-path from $x$ are parallel, and traverse same objects.\\
\hline
$inverse_{x,f,b,y}$ &
$reach_{x,f,y} \land \negFormula{x}{(\Edge{f} . \neg y)^*}{inv_{f,b}}$ \\&
doubly-linked lists between two variables $x$ and $y$ \\&
with $f$ and $b$ as forward and backward edges.\\
\hline
$tree_{root,r,l}$ & $\negFormula{root}{(\Edge{l} | \Edge{r})^*}{(uns_{l,r} \land
uns_{l} \land uns_{r})}
\land
\neg( \posFormula{root}{(\Edge{l} | \Edge{r})^*}{root} )$ \\
& tree rooted at $root$.\\
\hline
\end{tabular}
\end{center}
\caption{\label{Ta:Properties}%
Properties of data-structures expressed in $\LRPtwo$.}
\end{table*}
Using the patterns in \tableref{Pattern}, \tableref{Properties} defines
some interesting properties of data-structures using $\LRPtwo$.
The formula $reach_{x,f,y}$ means that the object pointed-to by the
program variable $y$ is reachable from the object pointed-to by the program variable $x$
by following an access path of $f$ field pointers.
We can also use it with $null$ in the place of $y$. For example,
the formula $reach_{x,f,null}$ describes a (possibly empty) linked-list pointed-to by $x$.
Note that it implies that the list is acyclic, because $null$ is always a
``sink'' node in a well-formed heap.
We can also express that there are no incoming $f$-edges into the list pointed to by $x$,
by conjoining the previous formula with $unshared_{x,f}$.
Alternatively, we can specify that $x$ is located on a cycle of $f$-edges: $cyclic_{x,f}$.
Disjointness can be expressed by the formula $disjoint_{x,f,y,g}$
that uses both forward and backward traversal of edges in the routing expression.
For example, we can express that the linked list pointed to by $x$ is disjoint from the linked-list pointed to by $y$,
using the formula $disjoint_{x,f,y,f}$.
Disjointness of data-structures is important for parallelization (e.g., see \cite{kn:HHN92}).

The last two examples in \tableref{Properties} specify data-structures with multiple fields.
The formula $inverse_{x,f,b,y}$ describes a doubly-linked with variables $x$ and $y$ pointing to the
head and the tail of the list, respectively.
First, it guarantees the existence of an $f$-path.
Next, it uses the pattern $inv_{f,b}$ to express that if there is an $f$-edge from one node to another,
then there is a $b$-edge in the opposite direction.
This pattern is applied to all nodes on the $f$-path that starts from $x$ and that does not visit $y$,
expressed using the test ``$\neg y$'' in the routing expression.
The formula $tree_{root,r,l}$ describes a binary tree.
The first part requires that the nodes reachable from the root (by following any path of $l$ and $r$ fields)
be not heap-shared. The second part prevents edges from pointing back to the root of the tree
by forbidding the root to participate in a cycle.


\newcommand{\verify}{\it VC}
\newcommand{\cond}{\it c}
\newcommand{\trans}{\it t}
\newcommand{\prsvh}[3]{{\it php}^{#1}_{#2}[#3]}
\newcommand{\prsvha}[2]{{\it phpa}^{#1}_{#2}}
\newcommand{\prsvst}[3]{{\it pst}^{#1}_{#2}[#3]}
\newcommand{\prsvsta}[2]{{\it psta}^{#1}_{#2}}

\subsection{Expressing Verification Conditions}

\def\ttlbrace{\char"7B}
\def\ttrbrace{\char"7D}
The \texttt{reverse} procedure  shown in \figref{rev} performs in-place reversal of a singly-linked list.
This procedure is interesting because it destructively updates the list and requires two fields
to express partial correctness. Moreover, it manipulates linked lists in which each list node
can be pointed-to from the outside.
In this section, we show that the verification conditions for the procedure \texttt{reverse}
can be expressed in $\LRPtwo$.
If the verification conditions are valid, then the program is partially correct with respect to the specification.
The validity of the verification conditions can be checked automatically because the logic $\LRPtwo$ is decidable,
as shown in the next section.
In \cite{tr05:VC}, we show how to automatically generate verification conditions in $\LRPtwo$
for arbitrary procedures that are annotated with preconditions, postconditions, and loop invariants in $\LRPtwo$.
\begin{figure}
\begin{center}
\begin{alltt}
\begin{tabbing}
No\=de reverse(Node x)\+\ttlbrace{}\\
L0: Node y = NULL;\\
L1: wh\=ile (x != NULL)\ttlbrace{}\\
L2: \> Node t = x->n;\\
L3: \> x->n = y;\\
L4: \> y = x;\\
L5: \> x = t;\\
L6: \ttrbrace{}\\
L7: return y;\-\\
\ttrbrace{}
\end{tabbing}
\end{alltt}
\end{center}
\caption{\label{Fig:rev} Reverse.}
\end{figure}

Notice that in this section we assume that all graphs denote valid stores,
i.e., satisfy~\equref{WF}.
The precondition requires that $x$ point to an acyclic list, on entry to the procedure.
We use the symbols $x^0$ and $n^0$ to record the values of the variable $x$
and the $n$-field on entry to the procedure.
\[
\begin{array}{c}
pre \eqdef \posFormula{x^0}{(\Edge{n^0})^*}{null^0}
\end{array}
\]
The postcondition ensures that the result is an acyclic list pointed-to by $y$.
Most importantly, it ensures that each edge of the original list is reversed in the
returned list, which is expressed in a similar way to a doubly-linked list, using $inverse$ formula.
We use the relation symbols $y^7$ and $n^7$ to refer to the values on exit.
\[
\begin{array}{c}
post \eqdef \posFormula{y^7}{(\Edge{n^7})^*}{null^7} \land
inverse_{x^0, n^0, n^7, y^7}
\end{array}
\]
The loop invariant $\phi$ shown below
relates the heap on entry to the procedure
to the heap at the beginning of each loop iteration (label \texttt{L1}).
First, we require that the part of the list reachable from $x$ be the
same as it was on entry to \texttt{reverse}.
Second, the list reachable from $y$ is reversed from its initial state.
Finally, the only original edge outgoing of $y$ is to $x$.
\[
\phi \eqdef  same_{x^1,n^0,n^1} \land inverse_{x^0,n^0,n^1,y^1}
 \land \posFormula{x^0}{\Edge{n^0}}{y^1}
\]
Note that the postcondition uses two binary relations, $n^0$ and $n^7$,
and also the loop invariant uses two binary relations, $n^0$ and $n^1$.
This illustrates that reasoning about singly-linked lists requires
more than one binary relation.

The verification condition of {\tt reverse} consists of two parts, $VC_{loop}$ and $VC$, explained below.

The formula $VC_{loop}$ expresses the fact that $\phi$ is indeed a loop invariant.
To express it in our logic, we use several copies of the vocabulary, one for each program point.
Different copies of the relation symbol $n$ in the graph
model values of the field $n$ at different program points.
Similarly, for constants. For example,
\figref{revLoop} shows a graph that satisfies the formula $VC_{loop}$ below.
It models the heap at the end of some loop iteration of \texttt{reverse}.
The superscripts of the symbol names denote the corresponding program points.
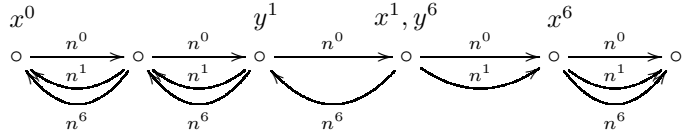
\begin{figure}
\begin{center}
$\xymatrix@R2pt@C30pt{
~~x^0 & & ~~y^1 & x^1,y^6 & ~~x^6
\\
\circ \ar[r]^{n^0} &
\circ \ar[r]^{n^0} \ar@/^1pc/[l]_{n^1} \ar@/^1.5pc/[l]^{n^6} &
\circ \ar[r]^{n^0} \ar@/^1pc/[l]_{n^1} \ar@/^1.5pc/[l]^{n^6} &
\circ \ar[r]^{n^0} \ar@/_1pc/[r]^{n^1}\ar@/^1.5pc/[l]^{n^6} &
\circ \ar[r]^{n^0} \ar@/_1pc/[r]^{n^1}\ar@/_1.5pc/[r]_{n^6} &
\circ \\
}$
\end{center}
\caption{\label{Fig:revLoop} An example graph that satisfies the $VC_{loop}$ formula for \texttt{reverse}.}
\end{figure}

To show that the loop invariant $\phi$ is maintained after executing the loop body, we assume that
the loop condition and the loop invariant hold at the beginning of the iteration,
and show that the loop body was executed without performing a null-dereference,
and the loop invariant holds at the end of the loop body:
\[
\begin{array}{rlr}
VC_{loop} \eqdef
& (x \neq null) & \mbox{loop is entered} \\
& \land \phi & \mbox{loop invariant holds on loop head}\\
& \land (y^6 = x^1) \land \posFormula{x^1}{n^1}{x^6} \land \posFormula{x^1}{n^6}{y^1} & \mbox{loop body}\\
& \land same_{y^1, n^1, n^6} \land same_{x^1,n^1,n^6} & \mbox{rest of the heap remains unchanged}\\
\implies
& (x^1 \neq null) & \mbox{no null-derefernce in the body}\\
& \land \phi^6 & \mbox{loop invariant after executing loop body}\\
\end{array}
\]
Here, $\phi^6$ denotes the loop-invariant formula $\phi$ after executing the loop body
(label \texttt{L6}), i.e., replacing all occurrences of $x^1$, $y^1$ and $n^1$
in $\phi$ by $x^6$, $y^6$ and $n^6$,
respectively. The formula $VC_{loop}$ defines a relation between three states:
on entry to the procedure, at the beginning of a loop iteration and at the end of a loop iteration.

The formula $VC$ expresses the fact that if the precondition holds
and the execution reaches procedure's exit
(i.e., the loop is not entered because the loop condition does not hold),
the postcondition holds on exit: $VC \eqdef pre \land (x^1 = null) \implies post$.

\section{Decidability of $\LRPtwo$}\label{Sec:decidable}

In this section, we show that $\LRPtwo$ is decidable for validity and satisfiability.
Since $\LRPtwo$ is closed under negation, it is sufficient to show that it is decidable for satisfiability.

The satisfiability problem for $\LRPtwo$ is decidable.
The proof proceeds as follows:
\begin{enumerate}
\item Every formula $\phi \in \LRPtwo$ can be
translated into an equi-satisfiable normal-form formula that is a disjunction of
formulas in $\NLBPtwo$ (\defref{NormalFormLBStwo}
and \theref{LBStwoTranslation}).
It is sufficient to show that the satisfiability of $\NLBPtwo$ is decidable.
\item
Define a class of simple graphs $\Ayah{k}$, for which the Gaifman graph is a tree with at
most $k$ additional edges (\defref{Ayahk}).
\item
Show that if formula $\phi  \in \NLBPtwo$ has a model, $\phi$ has a model in $\Ayah{k}$,
where $k$ is linear in the size of the formula $\phi$ (\theref{AyahModel}).
This is the main part of the proof.
\item
Translate formula $\phi  \in \NLBPtwo$ into an equivalent MSO formula\TrOnly{ (\lemref{TrMSO})}.
\item
Show that the satisfiability of MSO logic over $\Ayah{k}$ is decidable,
by reduction to MSO on trees \cite{Rabin}\TrOnly{ (\lemref{TrMona})}.
We could have also
shown decidability using the fact that the tree width of all graphs in $\Ayah{k}$
is bounded by $k$,
and that MSO over graphs with bounded tree width is decidable~\cite{Courcelle89,ALS91,Se92}.
\end{enumerate}

\begin{definition}\label{def:NormalFormLBStwo}\begin{Name}Normal-Form Formulas\end{Name}
A formula in $\NLBPtwo$ is a conjunction of reachability constraints
of the form $\posFormula{c_1}{R}{c_2}$ and $\negFormula{c}{R}{p}$, where
$p$ is one of the patterns allowed in $\LRPtwo$ (\defref{LBStwoPatterns}).
A normal-form formula is a disjunction of $\NLBPtwo$ formulas.
\end{definition}

\begin{theorem}\label{the:LBStwoTranslation}
There is a computable translation from $\LRPtwo$ to a disjunction of formulas in $\NLBPtwo$ that preserves
satisfiability.
\end{theorem}

\subsubsection{Ayah Graphs}
We define a notion of a simple tree-like directed graph, called Ayah graph.

Let $\gaifman{S}$ denote the \emph{Gaifman graph} of the graph $S$, i.e.,
an undirected graph obtained from $S$ by removing node labels, edge labels, and edge directions
(and parallel edges).
The \emph{distance} between nodes $v_1$ and $v_2$ in $S$
is the number of edges on the shortest path between $v_1$ and $v_2$ in $\gaifman{S}$.
An undirected graph $B$ is in $\ktree$ if removing self loops and at
most $k$ additional edges from $B$ results in an acyclic graph.

\begin{definition}\label{def:Ayahk}
For $k \geq 0$, an Ayah graph of $k$ is a graph $S$ for which the Gaifman graph is in $\ktree$:
$\Ayah{k} = \{ S | \gaifman{S} \in \ktree \}$.
\end{definition}

Let $\phi \in \NLBPtwo$ be of the form
$\phi_{\Diamond} \land \phi_{\Box} \land \phi_{=} \land \phi_{\rightarrow}$,
where $\phi_{\diamond}$ is a conjunction of
constraints of the form $\posFormula{c_1}{R}{c_2}$,
$\phi_{\Box}$ is a conjunction of reachability constraints with negative patterns,
$\phi_{=}$ is a conjunction of reachability constraints with  equality patterns,
and $\phi_{\rightarrow}$ is a conjunction of reachability constraints with edge patterns.

\begin{theorem}\label{the:AyahModel}
If $\phi \in \NLBPtwo$ is satisfiable, then $\phi$ is satisfiable by a graph in $\Ayah{k}$,
where $k = 2 \times n \times |C| \times m$,
$m$ is the number of constraints in $\phi_{\Diamond}$,
$|C|$ is the number of constants in the vocabulary, and
for every regular expression that appears in $\phi_{\Diamond}$ there
is an equivalent automaton with at most $n$ states.
\begin{Sketch}
Let $S$ be a model of $\phi$ : $S \models \phi$.
We construct a graph $S'$ from $S$ and show that $S' \models \phi$
\TrOnly{(\lemref{resModel})}
and $S' \in \Ayah{k}$\TrOnly{ (\lemref{resAyah})}.
The construction uses the following operations on graphs.

\paragraph{Witness Splitting}
A witness $W$ for a formula $\posFormula{c_1}{R}{c_2}$ in $\NLBPtwo$ in a graph $S$ is a path in $S$,
labelled with a word $w \in L(R)$, from the node labelled with $c_1$ to the node labelled with $c_2$.
Note that the nodes and edges on a witness path for $R$ need not be distinct.
Using $W$, we construct a graph $W'$ that consists of a path, labelled with $w$, that starts at the node labelled by $c_1$
and ends at the node labelled by $c_2$.
Intuitively, we duplicate a node of $W$ each time the witness path for $R$ traverses it,
unless the node is marked with a constant.
As a result, all shared nodes in $W'$ are labelled with constants.
Also, every cycle contains a node labelled with a constant.
By construction, we get that $W' \models \posFormula{c_1}{R}{c_2}$.
We say that $W'$ is the result of \emph{splitting} the witness $W$.
%

Finally, we say that $W$ is the \emph{shortest witness} for $\posFormula{c_1}{R}{c_2}$
if any other witness path for $\posFormula{c_1}{R}{c_2}$ is at least as long as $W$.
The result of splitting the shortest witness is a graph in $\Ayah{k}$, where
$k = 2 \times n \times |C|$: to break all cycles
it is sufficient to remove all the edges
adjacent to nodes labelled with constants,
and a node labelled with a constant is visited at most $n$ times.
(If a node is visited more than once in the same state of the automaton, the path can be
shortened.)

\begin{changebar}
\paragraph{Merge Operation}
Merging two nodes in a graph is defined in the usual way by gluing these nodes.
Let $p(v_0) \eqdef N(v_0,v_1,v_2) \implies (v_1 = v_2)$ be an equality pattern.
If a graph violates a reachability constraint \negFormula{c}{R}{p},
we can assign nodes $n_0$, $n_1$, and $n_2$ to $v_0$, $v_1$, and $v_2$, respectively,
such that there is a $R$-path from $c$ to $v_0$, $N(n_0,n_1,n_2)$ holds, and
$n_1$ and $n_2$ are distinct nodes.
In this case, we say that \emph{merge operation of $n_1$ and $n_2$
is enabled} (by \negFormula{c}{R}{p}).
The nodes $n_1$ and $n_2$ can be merge to discharge this assignment
(other merge operations might still be enabled after merging $n_1$ and $n_2$).

\paragraph{Edge-Addition Operation}
Let $p(v_0) \eqdef N(v_0, v_1, v_2) \implies v_1 \Edge{f} v_2$ be an edge pattern.
If a graph violates a reachability constraint \negFormula{c}{R}{p},
we can assign nodes $n_0$, $n_1$, and $n_2$ to $v_0$, $v_1$, and $v_2$, respectively,
such that there is a $R$-path from $c$ to $v_0$, $N(n_0,n_1,n_2)$ holds, and
there is no $f$-edge from $n_1$ to $n_2$.
In this case, we say that \emph{edge-operation operation is enabled} (by \negFormula{c}{R}{p}).
We can add an $f$-edge from $n_1$ and $n_2$ to discharge this assignment.
\end{changebar}

%

The following lemma is the key observation of this proof.
\begin{lemma}\label{lem:operationPresAyah}
The class of $\Ayah{k}$ graphs is closed under merge operations of nodes in distance at
most two and edge-addition operations at distance one.
\begin{Sketch}
\begin{changebar}
If an edge is added in parallel to an existing one (distance one),
it does not affect the Gaifman graph, thus $\Ayah{k}$ is closed under edge-addition.
The proof that $\Ayah{k}$ is closed under merge operations is more subtle~\cite{tr05:LRP}.
\end{changebar}
\end{Sketch}
\end{lemma}
In particular, the class $\Ayah{k}$ is closed under
the merge and edge-addition operations forced by $\LRPtwo$ formulas.
This is the only place in our proof where we use the distance restriction
of $\LRPtwo$ patterns.


Given a graph $S$ that satisfies $\phi$, we construct the graph $S'$ as follows:
\begin{enumerate}
\item
For each constraint $i$ in $\phi_{\diamond}$,
    identify the shortest witness $W_i$ in $S$.
    Let $W'_i$ be the result of splitting the witness $W_i$.
\item The graph $S_0$ is a union of all $W'_i$'s, in which the
    nodes labelled with the (syntactically) same constants are merged.
\item
    Apply all enabled merge operations and all enabled edge-addition operations in any order,
    producing a sequence of distinct graphs $S_0, S_1, \ldots, S_r$,
    until $S_m$ has no enabled operations.
\item The result $S' = S_r$.
\end{enumerate}
The process described above terminates after a finite number of steps,
because in each step either the number of nodes in the graph is decreased (by merge operations)
or the number of edges is increased (by edge-addition operations).

The proof proceeds by induction on the process described above.
Initially, $S_0$ is in $\Ayah{k}$. By \lemref{operationPresAyah},
all $S_i$ created in the third step of the construction above
are in $\Ayah{k}$; in particular, $S' \in \Ayah{k}$.

By construction of $S_0$, it contains a witness for each constraint in $\phi_{\Diamond}$,
and merge and edge-addition operations preserve the witnesses,
thus $S'$ satisfies $\phi_{\Diamond}$.
Moreover, $S_0$ satisfies all constraints in $\phi_{\Box}$.
We show that merge and edge-addition operations applied in the construction
preserve $\phi_{\Box}$ constraints,
thus $S'$ satisfies $\phi_{\Box}$.
The process above terminates when no merge and edge-addition operations are enabled,
that is, $S'$ satisfies $\phi_{=} \land \phi_{\rightarrow}$.
Thus, $S'$ satisfies $\phi$.

\SubOnly{The full proof is available at~\cite{tr05:LRP}.}
\end{Sketch}
\end{theorem}

\subsection{Complexity}

We proved decidability by reduction to MSO on trees, which allows us to
decide $\LRPtwo$ formulas using MONA decision procedure~\cite{KlaEtAl:Mona}.
Alternatively, a decision procedure for $\LRPtwo$ can directly construct a tree automaton
from a normal-form formula, and can then check emptiness of the automaton.
The worst case complexity of the satisfiability problem of $\LRPtwo$ formulas
is at least doubly-exponential, but it remains elementary
(in contrast to MSO on trees, which is non-elementary);
we are investigating tighter upper and lower bounds.
The complexity depends on the bound $k$ of $\Ayah{k}$ models, according to \theref{AyahModel}.
If the routing expressions do not contain constant symbols, then the bound $k$
does not depend on the routing expressions: it depends only
on the number of reachability constraints of the form
$\posFormula{c_1}{R}{c_2}$.
The $\LRPtwo$ formulas that come up in practice are well-structured,
and we hope to achieve a reasonable performance.


\section{Limitations and Further Extensions}\label{Sec:LimitExtend}

Despite the fact that $\LRPtwo$ is useful, there are interesting program properties
that cannot be expressed.
For example, transitivity of a binary relation,
that can be used, e.g., to express partial orders,
is naturally expressible in $\LRP$, but not in $\LRPtwo$.
Also, the property that a general graph is a tree in which each node has a pointer back to the root
is expressible in $\LRP$, but not in $\LRPtwo$.
Notice that the property is non-trivial
because we are operating on general graphs, and not just trees.
Operating on general graphs allows us to verify that the data-structure invariant
is reestablished after a sequence of low-level mutations that
temporarily violate the invariant data-structure.

There are of course interesting properties that are beyond $\LRP$,
such as the property that a general graph is a tree in which
every leaf has a pointer to the root of a tree.

In the future, we plan to generalize $\LRPtwo$ while maintaining decidability, perhaps beyond $\LRP$.
We are encouraged by the fact that the proof of decidability in \secref{decidable}
holds ``as is'' for many useful extensions.
For example, we can generalize the patterns to allow neighborhood formulas
with disjunctions and negations of unary relations.
In fact, more complex patterns can be used,
as long as they do not violate the $\Ayah{k}$ property.
For example, we can define trees rooted at $x$ with parent pointer $b$ from every tree node to its parent by
$tree_{x,r,l,b} \land  \textbf{let } p(v_0) \eqdef
((v_1 \Edge{l} v_0) \lor (v_1 \Edge{r} v_0)) \implies (v_0 \Edge{b} v_1)
\textbf{in }
\negFormula{x}{(\Edge{l} | \Edge{r})^*}{(det_{b} \land p)}
$.
The extended logic remains decidable, because the pattern $p$ adds edges only
in parallel to the existing ones.

Currently, reachability constraints describe paths that start from nodes labelled by constants.
We can show that the logic remains decidable when
reachability constraints are generalized to
describe paths that start from any node that satisfies a quantifier-free
\emph{positive} formula $\theta$:
$\forall v, w_0, \ldots, w_m, v_0, \ldots, v_n . R(v, v_0) \land \theta(v, w_0, \ldots, w_m) \implies ( N(v_0, \ldots, v_n) \implies
\psi(v_0, \ldots, v_n) )$.



\section{Related Work}\label{Sec:related}

There are several works on logic-based frameworks for reasoning about graph/heap structures.
We mention here the ones which are, as far as we know, the closest to ours.

The logic $\LRP$ can be seen as a fragment of the first-order logic over graph structures
with transitive closure (TC logic \cite{Immerman87}).
It is well known that TC is undecidable, and that this fact holds even when transitive closure
is added to simple fragments of FO such as the decidable fragment $L^2$
of formulas with two variables  \cite{Mort75,GKV,GOR99}.

It can be seen that our logics $\LRP$ and $\LRPtwo$ are both uncomparable with $L^2$ + TC.
Indeed, in $\LRP$ no alternation between universal and existential quantification is allowed.
On the other hand, $\LRPtwo$ allows us to express patterns (e.g., heap sharing) that
require more than two variables (see Table \ref{Ta:Pattern}, Section \ref{Sec:express}).

In \cite{ESOP:BRS99}, decidable logic  $L_r$ (which can also be seen as a fragment of TC) is introduced.
The logics $\LRP$ and $\LRPtwo$ generalize $L_r$, which is in fact the fragment of these
logics where only two fixed patterns are allowed: equality to a program variable and heap sharing.

In \cite{CSL:eadtc,BalabanPnueli:vmcai05,ShuvenduShaz:popl06,BCOH04} other decidable logics are defined,
but their expressive power is rather limited
w.r.t. $\LRPtwo$ since they allow at most one binary relation symbol (modelling linked data-structures
with 1-selector).
For instance, the logic of \cite{CSL:eadtc} does not allow us to express the reversal of a list.
Concerning the class of 1-selector linked data-structures,
\cite{BI05} provides a decision procedure for a logic with reachability constraints and arithmetical
constraints on lengths of segments in the structure.
It is not clear how the proposed techniques can be generalized to larger classes of graphs.
Other decidable logics \cite{BozgaIosifLakhnech:sas04,KuncakRinard04GeneralizedRecordsRoleLogic}
are restricted in the sharing patterns and the reachability they can describe.

Other works in the literature consider extensions of the first-order logic with fixpoint operators.
Such an extension is again undecidable in general but the introduction of the notion
of (loosely) guarded quantification allows one to obtain decidable fragments such as
$\mu GF$ (or $\mu LGF$)
(Guarded Fragment with least and greater fixpoint operators) \cite{GW99,Gr02}.
Similarly to our logics, the logic $\mu GF$ (and also $\mu LGF$) has the tree model property:
every satisfiable formula has a model of bounded tree width.
However, guarded fixpoint logics are incomparable with $\LRP$ and $\LRPtwo$.
For instance, the $\LRPtwo$ pattern $det_f$ that requires determinism of $f$-field,
is not a (loosely) guarded formula.

The PALE system \cite{PLDI:MS01} uses an extension of the monadic second order logic on trees
as a specification language.
The considered linked data structures are those that can be defined as {\em graph types} \cite{KS93}.
Basically, they are graphs that can be defined as
trees augmented by a set of edges defined using routing expressions
(regular expressions) defining paths in the (undirected structure of the) tree.
$\LRPtwo$ allows us to reason naturally about arbitrary graphs without
limitation to tree-like structures.
Moreover, as we show in Section \ref{Sec:express}, our logical framework allows us
to express postconditions and
loop invariants that relate the input and the output state. For instance,
even in the case of singly-linked lists, our framework allows us to express properties
that cannot be expressed in the PALE framework:
in the list reversal example of Section \ref{Sec:express},
we show that the output list is precisely the reversed input list,
whereas in the PALE approach, one can only establish that the output is
a list that is the permutation of the input list.

In~\cite{cav04:IRRSY}, we tried to employ a decision procedure for MSO on trees to reason
about reachability.
However, this places a heavy burden on the specifier to prove that the data-structures
in the program can be simulated using trees.
The current paper eliminated this burden by defining syntactic
restrictions on the formulas and showing a general reduction theorem.

Other approaches in the literature use undecidable formalisms such as
\cite{kn:HHN92}, which provides a natural and expressive language,
but does not allow for automatic property checking.

Separation logic has been introduced recently as a formalism for reasoning about heap
structures \cite{Rey02}.
The general logic is undecidable \cite{CYOH01} but there are few works showing decidable fragments
\cite{CYOH01,BCOH04}.
One of the fragments is propositional separation logic where quantification is
forbidden \cite{CYOH01,CGH05}
and therefore seems to be incomparable with our logic. The fragment defined
in \cite{BCOH04} allows one to reason only about singly-linked lists with explicit sharing.
In fact, the fragment considered in \cite{BCOH04} can be translated to $\LRPtwo$,
and therefore, entailment problems as stated in \cite{BCOH04} can be solved as implication problems
in $\LRPtwo$.

\section{Conclusions}

Defining decidable fragments of first order logic with
transitive closure that are useful for program verification is a difficult task (e.g.,~\cite{CSL:eadtc}).
In this paper, we demonstrated that this is possible by combining three principles:
(i)~allowing arbitrary boolean combinations of the reachability constraints,
which are closed formulas without quantifier alternations,
(ii)~defining reachability using regular expressions denoting
pointer access paths (not) reaching a certain pattern, and
(iii)~syntactically limiting the way patterns are formed.
Extensions of the patterns that allow larger distances between
nodes in the pattern either break our proof of decidability or are directly undecidable.

The decidability result presented in this paper improves the state-of-the-art significantly.
In contrast to \cite{CSL:eadtc,BalabanPnueli:vmcai05,ShuvenduShaz:popl06,BCOH04},
$\LRP$ allows several binary
relations. This provides a natural way to
(i)~specify invariants for data-structures with multiple fields (e.g., trees,
doubly-linked lists),
(ii)~specify post-condition for procedures that mutate pointer fields of
data-structures, by expressing the relationships between fields before and
after the procedure (e.g., list reversal, which is beyond the scope of PALE),
(iii)~express verification conditions using a copy of the vocabulary for
each program location.

We are encouraged by the expressiveness of this simple logic and plan to explore its usage
for program verification and abstract interpretation.

\bibliography{LBS}

\begin{thebibliography}{10}

\bibitem{ALS91}
S.~Arnborg, J.~Lagergren, and D.~Seese.
\newblock Easy problems for tree-decomposable graphs.
\newblock {\em J. Algorithms}, 12(2):308--340, 1991.

\bibitem{BalabanPnueli:vmcai05}
I.~Balaban, A.~Pnueli, and L.~D. Zuck.
\newblock Shape analysis by predicate abstraction.
\newblock In {\em VMCAI}, pages 164--180, 2005.

\bibitem{ESOP:BRS99}
M.~Benedikt, T.~Reps, and M.~Sagiv.
\newblock A decidable logic for describing linked data structures.
\newblock In {\em European Symp. On Programming}, pages 2--19, March 1999.

\bibitem{BCOH04}
J.~Berdine, C.~Calcagno, and P.~O'Hearn.
\newblock {A Decidable Fragment of Separation Logic}.
\newblock In {\em FSTTCS'04}. LNCS 3328, 2004.

\bibitem{BHMV05}
A.~Bouajjani, P.~Habermehl, P.Moro, and T.~Vojnar.
\newblock {Verifying Programs with Dynamic 1-Selector-Linked Structures in
  Regular Model Checking}.
\newblock In {\em Proc. of TACAS '05}, volume 3440 of {\em LNCS}. Springer,
  2005.

\bibitem{BI05}
M.~Bozga and R.~Iosif.
\newblock {Quantitative Verification of Programs with Lists}.
\newblock In {\em VISSAS intern. workshop}. IOS Press, 2005.

\bibitem{BozgaIosifLakhnech:sas04}
M.~Bozga, R.~Iosif, and Y.~Lakhnech.
\newblock On logics of aliasing.
\newblock In {\em Static Analysis Symp.}, pages 344--360, 2004.

\bibitem{JML}
L.~Burdy, Y.~Cheon, D.~Cok, M.~Ernst, J.~Kiniry, G.~T. Leavens, K.~R.~M. Leino,
  and E.~Poll.
\newblock An overview of jml tools and applications.
\newblock {\em Int. J. on Software Tools for Technology Transfer},
  7(3):212--232, 2005.

\bibitem{CGH05}
C.~Calcagno, P.~Gardner, and M.~Hague.
\newblock {From Separation Logic to First-Order Logic}.
\newblock In {\em FOSSACS'05}. LNCS 3441, 2005.

\bibitem{CYOH01}
C.~Calcagno, H.~Yang, and P.~O'Hearn.
\newblock {Computability and Complexity Results for a Spatial Assertion
  Language for Data Structures}.
\newblock In {\em FSTTCS'01}. LNCS 2245, 2001.

\bibitem{Courcelle89}
B.~Courcelle.
\newblock The monadic second-order logic of graphs, ii: Infinite graphs of
  bounded width.
\newblock {\em Mathematical Systems Theory}, 21(4):187--221, 1989.

\bibitem{Gr02}
E.~Gr{\"a}del.
\newblock Guarded fixed point logic and the monadic theory of trees.
\newblock {\em Theoretical Computer Science}, 288:129--152, 2002.

\bibitem{GOR99}
E.~Gr{\"{a}}del, M.Otto, and E.Rosen.
\newblock Undecidability results on two-variable logics.
\newblock {\em Archive of Math. Logic}, 38:313–--354, 1999.

\bibitem{GW99}
E.~Gr{\"{a}}del and I.~Walukiewicz.
\newblock {Guarded Fixed Point Logic}.
\newblock In {\em LICS'99}. IEEE, 1999.

\bibitem{GKV}
E.~Graedel, P.~Kolaitis, and M.~Vardi.
\newblock On the decision problem for two variable logic.
\newblock {\em Bulletin of Symbolic Logic}, 1997.

\bibitem{kn:Hendren}
L.~Hendren.
\newblock {\em Parallelizing Programs with Recursive Data Structures}.
\newblock PhD thesis, Cornell Univ., Ithaca, NY, Jan 1990.

\bibitem{kn:HHN92}
L.~Hendren, J.~Hummel, and A.~Nicolau.
\newblock Abstractions for recursive pointer data structures: {I}mproving the
  analysis and the transformation of imperative programs.
\newblock In {\em SIGPLAN Conf. on Prog. Lang. Design and Impl.}, pages
  249--260, New York, NY, June 1992. ACM Press.

\bibitem{KlaEtAl:Mona}
J.G. Henriksen, J.~Jensen, M.~J{\o}rgensen, N.~Klarlund, B.~Paige, T.~Rauhe,
  and A.~Sandholm.
\newblock Mona: Monadic second-order logic in practice.
\newblock In {\em {TACAS}}, 1995.

\bibitem{kn:Hoare75}
C.A.R. Hoare.
\newblock Recursive data structures.
\newblock {\em Int. J. of Comp. and Inf. Sci.}, 4(2):105--132, 1975.

\bibitem{Immerman87}
N.~Immerman.
\newblock Languages that capture complexity classes.
\newblock {\em SIAM Journal of Computing}, 16:760--778, 1987.

\bibitem{CSL:eadtc}
N.~Immerman, A.~Rabinovich, T.~Reps, M.~Sagiv, and G.~Yorsh.
\newblock The boundery between decidability and undecidability of transitive
  closure logics.
\newblock In {\em CSL}, 2004.

\bibitem{cav04:IRRSY}
N.~Immerman, A.~Rabinovich, T.~Reps, M.~Sagiv, and G.~Yorsh.
\newblock Verification via structure simulation.
\newblock In {\em CAV}, 2004.

\bibitem{POPL:OI01}
S.~S. Ishtiaq and P.~W. O'Hearn.
\newblock Bi as an assertion language for mutable data structures.
\newblock In {\em POPL}, pages 14--26, 2001.

\bibitem{KS93}
N.~Klarlund and M.~I. Schwartzbach.
\newblock {Graph Types}.
\newblock In {\em POPL'93}. ACM, 1993.

\bibitem{KuncakRinard04GeneralizedRecordsRoleLogic}
V.~Kuncak and M.~Rinard.
\newblock Generalized records and spatial conjunction in role logic.
\newblock In {\em Static Analysis Symp.}, Verona, Italy, August 26--28 2004.

\bibitem{ShuvenduShaz:popl06}
S.~K. Lahiri and S.~Qadeer.
\newblock Verifying properties of well-founded linked lists.
\newblock In {\em Symp. on Princ. of Prog. Lang.}, 2006.
\newblock To appear.

\bibitem{SAS:LS00}
T.~Lev-Ami and M.~Sagiv.
\newblock {TVLA}: {A} system for implementing static analyses.
\newblock In {\em Static Analysis Symp.}, pages 280--301, 2000.

\bibitem{PLDI:MS01}
A.~M{\o}ller and M.I. Schwartzbach.
\newblock The pointer assertion logic engine.
\newblock In {\em SIGPLAN Conf. on Prog. Lang. Design and Impl.}, pages
  221--231, 2001.

\bibitem{Mort75}
M.~Mortimer.
\newblock On languages with two variables.
\newblock {\em Zeitschrift f\"{u}r Mathematische Logik und Grundlagen der
  Mathematik}, 21:135--140, 1975.

\bibitem{Rabin}
M.~Rabin.
\newblock Decidability of second-order theories and automata on infinite trees.
\newblock {\em Trans. Amer. Math. Soc.}, 141:1--35, 1969.

\bibitem{CAV:RSW04}
T.~Reps, M.~Sagiv, and R.~Wilhelm.
\newblock Static program analysis via 3-valued logic.
\newblock In {\em CAV}, pages 15--30, 2004.

\bibitem{Rey02}
J.~C. Reynolds.
\newblock {Separation Logic: A Logic for Shared Mutable Data Structures}.
\newblock In {\em LICS'02}. IEEE, 2002.

\bibitem{kn:SRW98}
M.~Sagiv, T.~Reps, and R.~Wilhelm.
\newblock Solving shape-analysis problems in languages with destructive
  updating.
\newblock {\em ACM Transactions on Programming Languages and Systems},
  20(1):1--50, January 1998.

\bibitem{TOPLAS:SRW02}
M.~Sagiv, T.~Reps, and R.~Wilhelm.
\newblock Parametric shape analysis via 3-valued logic.
\newblock {\em ACM Transactions on Programming Languages and Systems}, 2002.

\bibitem{Se92}
D.~Seese.
\newblock Interpretability and tree automata: A simple way to solve algorithmic
  problems on graphs closely related to trees.
\newblock In {\em Tree Automata and Languages}, pages 83--114. 1992.

\bibitem{tr05:LRP}
G.~Yorsh, M.~Sagiv, A.~Rabinovich, A.~Bouajjani, and A.~Meyer.
\newblock A logic of reachable patterns in linked data-structures.
\newblock Technical report, Tel Aviv University, 2005.
\newblock Available at ``www.cs.tau.ac.il/$\sim$gretay''.

\bibitem{tr05:VC}
G.~Yorsh, M.~Sagiv, A.~Rabinovich, A.~Bouajjani, and A.~Meyer.
\newblock Verification framework based on the logic of reachable patterns.
\newblock In preparation, 2005.

\end{thebibliography}
\bibliographystyle{plain}

\appendix
\TrOnly{

\section{Semantics of $\LRP$}

The semantics of $\LRP$ formulas is formally defined as follows.

\begin{Definition}\label{def:FormulaMeaning}
\textbf{An interpretation} for an $\LRP$ formula over $\voc = \B{C, U, F}$
is a labelled directed graph $G = \B{V^G, E^G, C^G, U^G}$
where:
(i)~$V^G$ is a set of nodes modelling the heap objects, 
(ii)~$E^G \colon F \to  \mathcal{P}(V^G \times V^G)$ are labelled edges,
(iii)~$C^G \colon C \to V^G$ maps constant symbols to their nodes,
and
(iv)~$U^G \colon U \to \mathcal{P}(V^G)$ maps unary relation symbols to the nodes in which they hold.

Consider a routing expression $R$ and $w \in L(R)$.
We say that \textbf{there is a path  labelled by $w$ from a node $v_1$ to a node $v_2$}
if one of the following conditions hold:
\begin{itemize}
\item
$v_1 = v_2$ and $w= \epsilon$,
\item
$v_1 = v_2$, $w = u$ for a unary relation symbol $u$ and $v_1 \in U^G(u)$,
\item
$v_1 = v_2$, $w = \neg u$ for a unary relation symbol $u$ and $v_1 \notin U^G(u)$,
\item
$v_1 = v_2$, $w = c$ for a constant $c$ and $C^G(c) = v_1$,
\item
$v_1 = v_2$, $w = \neg c$ for a constant $c$ and $C^G(c) \neq v_1$,
\item
$w = \Edge{f}$ for an edge $f \in F$ and $\B{v_1, v_2} \in E^G(f)$,
\item
$w = \BEdge{f}$ for an edge $f \in F$ and $\B{v_2, v_1} \in E^G(f)$,
\item~$w = w_1 . w_2$  and there exists a node $v_3$ such that
there is a path labelled by $w_1$ from $v_1$ to $v_3$  and there
exists a path labelled by $w_2$ from $v_3$ to $v_2$ .
\end{itemize}

We say that a node tuple in $G$ satisfies a pattern $p$
if it satisfies the quantifier-free formula
that defines $p$, according to the usual semantics of the first-order logic over graph structures.

Finally, we define a satisfaction relation $\models$ between a graph $G$
and $\LRP$ formulas as follows:
we say that $G$ satisfies a formula
$\phi = \pred{c}{R}{p}$
(and we write $G \models \phi$) if and only if
for every $w \in L(R)$ and for every node tuple $u_0, \ldots, u_n$ in $G$,
if there is a path labelled by $w$ from $c$ to $u_0$,
then the tuple $u_0, \ldots, u_n$, where $u_0$ used as a central node, satisfies $p$.
The meaning of Boolean connectives is defined in a standard way.
\end{Definition}

\section{Proof of Undecidability}

\begin{SThe}{the:Undecidable}
The satisfiability problem of  $\LRP$ formulas is undecidable.
\begin{Proof}
Given a Turing machine $M$, we construct a formula $\varphi_{M}$ of size $O(poly(|M|))$,
such that $\varphi_{M}$ is satisfiable if and only if the execution of $M$ eventually halts.

The idea is that each node in the graph that satisfies $\varphi_{M}$ describes a cell of a tape in
some configuration, with unary relation symbols encoding the symbol in each cell,
the location of the head and the current state.
The $n$-edges describe the sequence of cells in a configuration and a sequence of configurations.
The $b$-edges describe how the cell is changed from one configuration to the next.
The constant $c_1$ marks the node that describes the first cell of the tape in the first configuration,
the constant $c_2$ marks the node that describes the first cell in the second configuration,
and the constant $c_3$ marks the node that describes the last cell in the last configuration
(see sketch in \figref{model}).

Let $M = \B{\Gamma,Q,q_0,q_n,\delta}$ denote a Turing machine where
$\Gamma = \{ \sigma_0, \ldots, \sigma_k \} \cup \blank$ are the symbols of the tape of $M$,
$Q = q_0, \ldots, q_n$ is the set of states of $M$,
$q_0$ is the initial state, $q_n$ is the halting state (with $n > 0$),
and $\delta$ is a (deterministic) transition relation:
$\B{q_i, s, q_j, s', \{R,L,D\}} \in \delta$ means that in the state $q_i$ and reading the symbol $s$ on the tape,
the Turing machine writes $s'$ on the tape, goes to state $q_j$ and moves the head right $R$  or left $L$ or do not move $D$.

We need unary relation symbols $q_0, \ldots, q_n$, $\sigma_0, \ldots, \sigma_k$ to encode states, symbols and the location of the head on the tape.

\begin{enumerate}
\item There is $n$-path from $c_1$ to $c_2$: $\posFormula{c_1}{(\Edge{n})^*}{c_2}$
\item There is $n$-path from $c_2$ to $c_3$: $\posFormula{c_2}{(\Edge{n})^*}{c_3}$
\item There is $b$-edge from $c_1$ to $c_2$ : $\posFormula{c_1}{\Edge{b}}{c_2}$.
\item No $n$-edge exits $c_3$: $\negFormula{c_3}{\Edge{n}}{false}$.
\item For every node $v$ that is $n$-reachable from $c_1$, if there is a $b$-edge from $v$
to $u$, then there is a $b$-edge from the $n$-successor of $v$ to the
$n$-successor of $u$:
$\textbf{let } p(v) \eqdef (v \Edge{b} u) \land (v \Edge{n} v_1) \land (u \Edge{n} u_1)
\implies (v_1 \Edge{b} u_1)
\textbf{ in }~\negFormula{c_1}{(\Edge{n})^*}{p}$.
\item The $n$-edges and the $b$-edges reachable from $c_1$ are deterministic: $\negFormula{c_1}{(\Edge{n})^*}{(det_n \land det_b)}$.
\item Initial configuration encodes an empty tape (a sequence of blanks of some length)
with the head pointing to the first cell on the tape, in the state $q_0$.
Other cells between $c_1$ and $c_2$ do not have state constraints.
\[
\begin{array}{ll}
\textbf{let}~p(v) \eqdef true \implies  q_0(v)~\textbf{in}~\negFormula{c_1}{\epsilon}{p}\\
\textbf{let}~p(v) \eqdef true \implies  \blank(v)~\textbf{in}~\negFormula{c_1}{(\Edge{n}. \neg c_2)^*}{p}\\
\textbf{let}~p(v) \eqdef true \implies \Land_{q \in Q} \neg q(v)~\textbf{in}~\neg \negFormula{c_1}{(\Edge{n}. \neg c_2)^+}{p}\\
\end{array}
\]

\item The accepting state $q_n$ is reachable from $c_2$: $\posFormula{c_2}{(\Edge{n})^*}{q_n}$.

\item At least one symbol at each node,
at most one symbol at each node, and
at most one state at each node:
\[
\begin{array}{l}
\textbf{let }p(v) \eqdef true \implies \Lor_{\sigma \in \Gamma} \sigma(v)\textbf{ in }\negFormula{c_1}{(\Edge{n})^*}{p}\\
\textbf{let }p(v) \eqdef true \implies  \Land_{\sigma_i \neq \sigma_j \in \Gamma} \neg (\sigma_i(v) \land \sigma_j(v)) \textbf{ in }\negFormula{c_1}{(\Edge{n})^*}{p}\\
\textbf{let }p(v) \eqdef true \implies  \Land_{q_i \neq q_j \in Q} \neg (q_i(v) \land q_j(v)) \textbf{ in }\negFormula{c_1}{(\Edge{n})^*}{p}
\end{array}
\]

\item For each transition $\B{q_i, s, q_j, s', D} \in \delta$,
\[
\textbf{let }p(v) \eqdef (v \Edge{b} w) \implies (q_i(v) \land s(v) \implies (q_j(w) \land s'(v))) \textbf{ in }
\negFormula{c_1}{(\Edge{n})^*}{p}
\]
\item For each transition $\B{q_i, s, q_j, s', R\}} \in \delta$,
\[
\begin{array}{ll}
\textbf{let }p(v) \eqdef & (v \Edge{b} w) \land (v \Edge{n} v') \land (v' \Edge{b} w') \implies \\&
q_i(v) \land s(v) \implies  (q_j(w') \land s'(v) \land (\Land_{q \in Q} \neg q(v)))
\textbf{ in }\negFormula{c_1}{(\Edge{n})^*}{p}
\end{array}
\]
Similar for the case in which the head moves to the left.
\item States do not come ``out of the blue'' : for each $q_j \in Q$,
\[
\begin{array}{rl}
\textbf{let } p(w) \eqdef & (v \Edge{b} w) \land (w' \Edge{n} w) \land (w \Edge{n} w'') \land (v' \Edge{b} w')
\land (v'' \Edge{b} w'') \implies
s_j(w) \land q_j(w) \\
 & \implies \Lor_{\B{q_i, s_i, q_j, s_j, D} \in \delta} q_i(v) \land s_i(v)\\
 & \lor \Lor_{\B{q_i, s_i, q_j, s_j, R}  \in \delta} q_i(v') \land s_i(v')
\lor \Lor_{\B{q_i, s_i, q_j, s_j, L}  \in \delta} q_i(v'') \land s_i(v'')\\
\textbf{ in } & \negFormula{c_1}{(\Edge{n})^*}{p}
\end{array}
\]
\item Exactly one node in a given configuration is labelled by a state.
This holds in the initial configuration (see $7$), to ensure this in the following configurations, we require that
a transition does not cross configuration boundary.
$L$-transitions cannot be enabled on the first node in a configuration:
\[
\textbf{let }p(v) \eqdef true \implies \Land_{\B{q,s,q',s',L} \in \delta}
        \neg (q(v) \land s(v)) \textbf{ in } \negFormula{c_1}{(\Edge{b})^*}{p}.
\]
Similarly, $R$-transitions cannot be enabled on the last node in a configuration: for all $\B{q,s,q',s',R} \in \delta$,
\[
\textbf{let }p(v) \eqdef (w \Edge{n} v ) \implies \Land_{\B{q,s,q',s',R} \in \delta}
        \neg (q(w) \land s(w)) \textbf{ in } \negFormula{c_1}{(\Edge{b})^*}{p}.
\]
\item Symbols not pointed to by the head remain the same in the next configuration:
$\textbf{let }p(v) \eqdef (v \Edge{b} w) \implies
(s(v) \land \Land_{q \in Q} \neg q(v)) \implies s(w) \textbf{ in }\negFormula{c_1}{(\Edge{n})^*}{p}$.
\end{enumerate}
\end{Proof}
\end{SThe}


\section{Decidability of $\LRPtwo$}

In this section, we give the details of the proof of \theref{AyahModel}.
To simplify the presentation, we consider only $\LRPtwo$ formulas
in which the patterns do not contain
positive occurrences of unary relations and equalities that involve constants.
In the proofs, we point out where this assumption is essential, and hint on how to modify the proof.


\subsection{Normal Form of $\LRPtwo$-Formulas}

Every formula $\phi \in \LRPtwo$ can be translated into an equi-satisfiable normal-form formula.
A normal-form formula is a disjunction of $\NLBPtwo$ formulas,
where each formula in $\NLBPtwo$ is a conjunction of reachability constraints (\defref{NormalFormLBStwo}).
Note that while $\LRPtwo$ is closed under negation, $\NLBPtwo$ is not.

\begin{SThe}{the:LBStwoTranslation}
There is a computable translation $TR$ from $\LRPtwo$ to a disjunction of formulas in $\NLBPtwo$ that preserves
satisfiability.
\begin{Sketch}
For every formula $\phi \in \LRPtwo$ over $\voc$, $TR(\phi)$ is a disjunction of formulas in $\NLBPtwo$ over $\voc'$
such that $\phi$ is satisfiable if and only if $TR(\phi)$ is satisfiable.
The vocabulary $\voc'$ is an extension of $\voc$ with new constant symbols (introduced below)
The translation $TR(\phi)$ is defined as follows:
\begin{enumerate}
\item Transform $\phi$ into an equivalent formula $\phi'$ that is a
disjunction of conjunctions of reachability constraints $\negFormula{c}{R}{p}$ and their negations.

\item For every reachability constraint $\negFormula{c}{R}{p}$ that appears in $\phi'$ under negation,
if $p(v_0) = N(v_0, \ldots, v_n) \implies \psi(v_0, \ldots, v_n)$,
we introduce new constant symbols $c_0, \ldots, c_n$.
We can rewrite $\neg \negFormula{c}{R}{p}$ as a boolean combination of reachability constraints of the form
$\posFormula{c'}{R'}{c''}$ by replacing all variables with the corresponding constant symbols.
The result is $\posFormula{c}{R}{c_0} \land (N(c_0, \ldots, c_n) \implies \psi(c_0, \ldots, c_n))$,
where every edge formula $v_i \Edge{f} v_j$ that appears in $N$ or
$\psi$ becomes a simple reachability constraint $\posFormula{c_i}{\Edge{f}}{c_j}$.
Finally, transform the result into disjunctive normal form, denoted by $\phi''$.
The formula $\phi''$ is satisfiable if and only if $\phi'$ is satisfiable.

\end{enumerate}
Note that the first transformation preserves equivalence,
but the second
only preserve satisfiability.
The translation is applicable to the full $\LRP$ logic,
in which case the reachability constraints will contain arbitrary patterns.
\end{Sketch}
\end{SThe}

In the rest of the proof, let $\phi \in \NLBPtwo$
be of the form $\phi_{\Diamond} \land \phi_{\Box} \land \phi_{=} \land \phi_{\rightarrow}$,
where $\phi_{\diamond}$ is a conjunction of
reachability constraints of the form $\posFormula{c_1}{R}{c_2}$,
$\phi_{\Box}$ is a conjunction of reachability constraints with negative patterns,
$\phi_{=}$ is a conjunction of reachability constraints with  equality patterns,
and $\phi_{\rightarrow}$ is a conjunction of reachability constraints with edge patterns.

%
\subsubsection{Merge Operation}
Let $p(v_0) \eqdef N(v_0,v_1,v_2) \implies (v_1 = v_2)$ be an equality pattern.
If a graph violates a reachability constraint with pattern $p$,
we can assign distinct nodes to $v_1$ and $v_2$.
We say that a \emph{merge operation is enabled}, i.e., the nodes assigned to $v_1$ and $v_2$
can be merged to discharge this unsatisfying assignment to $p$
(other assignments might exist in the same graph).

Formally, the merge operation of $v_1$ and $v_2$ in $S$, which results in a graph $S'$,
is a transformation $m \colon S \to S'$.
The set of nodes of $S'$ is $S\setminus\{v_1,v_2\}\cup\{v_{12}\}$,
where $v_{12}$ is a new node.
Let  $m \colon S \to S'$ be defined as follows:
\[
m(v) = \left\{\begin{array}{ll}
v_{12} & \mbox{ if } v = v_1 \mbox{ or } v = v_2\\
v & \mbox{ otherwise}
\end{array}\right.
\]
The interpretation of constant and relation symbols in $S'$ is defined as follows:
\begin{enumerate}
\item  $c^{S'} = m (c^{S})$ for every constant symbol $c \in \voc$.
\item for every unary relation symbol $\sigma \in
\voc$, and for every $v\in S$, if $v$ is labelled with $\sigma$ in
$S$ then $m(v)$ is labelled with $\sigma$ in $S'$.
\item for every binary relation symbol $\sigma \in \voc$,
and every pair of nodes $v_1, v_2 \in S$, if there is an edge from
$v_1$ to $v_2$ labelled with $\sigma$ then there is an edge from
$m(v_1)$ to $m(v_2)$ in $S_2$ labelled with $\sigma$.
\end{enumerate}

\subsubsection{Witness Splitting}
A witness $W$ for a formula $\posFormula{c_1}{R}{c_2}$ in $\NLBPtwo$
in a graph $S$ is a path in $S$,
labelled with a word $w \in L(R)$, from the node labelled
with $c_1$ to the node labelled with $c_2$.
Note that the nodes and edges on a witness path for $R$ need not be distinct.
Using $W$, we construct a graph $W'$ that consists of a path,
also labelled with $w$, that starts at the node labelled by $c_1$
and ends at the node labelled by $c_2$.
Intuitively, we duplicate a node of $S$ each time the witness path $W$
traverses it, unless the node is marked with a constant.
As a result, all shared nodes in $W'$ are labelled with constants.
Also, every cycle contains a node labelled with a constant.
By construction, we get that $W' \models \posFormula{c_1}{R}{c_2}$.
We say that $W'$ is the result of \emph{splitting} the witness $W$.

Formally, the witness path $W$ is a sequence of nodes from $S$:
 $t_1, t_2, \ldots, t_r$, where $t_i \in S$.
For a node $t \in S$, let $C(t)$ denote the set of constant symbols that label the node $t$.
We define a mapping $d(t_i)$ as follows:
\[
d(t_i) \eqdef
\left\{\begin{array}{ll}
t_{C(t_i),0} & \mbox{if}~C(t_i)\neq \emptyset\\
t_{v,l} & \mbox{if}~t~\mbox{is the }l\mbox{-th occurrence of the node}~v\in S~\mbox{on the path}~W
\end{array}\right.
\]

$W'$ is a graph with nodes $d(t_1), \ldots, d(t_r)$.
If the witness path $W$ goes from $t_i$ to $t_{i+1}$ through
an edge labelled with $f_{i} \in F$, then there is an edge in $W'$
labelled with $f_{i}$ from $d(t_i)$ to $d(t_{i+1})$.
Note that $W'$ contains only edges traversed by the witness path.
For every constant symbol $\sigma \in C$ and node $t_i \in W$,
$d(t_i)$ is labelled with $\sigma$ in $W'$ if and only if $\sigma \in C(t_i)$.
For every unary relation symbol $\sigma \in U$ and node $t_i \in W$,
$d(t_i)$ is labelled with $\sigma$ in $W'$ if and only if $t_i$ is labelled with $\sigma$ in $S$.

Finally, we say that $W$ is the \emph{shortest witness} for $\posFormula{c_1}{R}{c_2}$
if any other witness path for $\posFormula{c_1}{R}{c_2}$ is at least as long as $W$.

\subsubsection{Homomorphism Preservation}

We give a slightly non-standard definition of homomorphism between graphs.
The homomorphism relation defined below preserves existence of edges
and both existence and absence of labels on nodes
(preserving absence of labels is non-standard).
We show that this homomorphism relation is preserved by $\NLBPtwo$ formulas,
and also by merging and splitting operations.
Simple proofs are omitted. 

\begin{Def}\begin{Name}Homomorphism\end{Name}\label{def:homomorphism}
Let $S_1$ and $S_2$ be graphs over the same vocabulary $\voc$.
A homomorphism from $S_1$ to $S_2$ is a mapping $h \colon S_1 \to S_2$ such that
\begin{enumerate}
\item for every constant symbol and unary relation symbol $\sigma \in \voc$,
and for every $v\in S_1$,
$v$ is labelled with $\sigma$ in $S_1$ if and only if $h(v)$ is labelled with $\sigma$ in $S_2$.
\item for every binary relation symbol $\sigma \in \voc$,
and every pair of nodes $v_1, v_2 \in S_1$,
if there is an edge $(v_1,v_2)$ labelled with $\sigma$, then there is an edge $(h(v_1), h(v_2))$ in $S_2$
labelled with $\sigma$.
\end{enumerate}
\end{Def}


\begin{Lemma}\label{lem:homoPres}
Let $h \colon S_1 \to S_2$ be a homomorphism.
If $\phi$ is a reachability constraint $\posFormula{c_1}{R}{c_2}$,
then if $S_1 \models \phi$ then $S_2 \models \phi$.
Dually, if $\phi$ is a reachability constraint  $\negFormula{c}{R}{p}$ in $\NLBPtwo$, then
if $S_2 \models \phi$ then $S_1 \models \phi$.
\end{Lemma}

\begin{Lemma}\label{lem:splittingHomo}
Let $W$ be a witness for $\posFormula{c_1}{R}{c_2}$ in $S$,
and let $W'$ be a splitting of $W$.
There is a homomorphism from $W'$ to $S$.
\end{Lemma}

\begin{Lemma}\label{lem:mergeHomo}
Assume that $f$ is a homomorphism  from $S_1$ to $S$ such
that $f(v_1)=f(v_2)$ and that $S_2$ is obtained by merging the nodes $v_1$ and $v_2$ in $S_1$.
There is a homomorphism from $S_2$ to $S$.
\end{Lemma}

\subsubsection{Properties of Ayah Graphs}

%
%
%
%

\begin{Lemma}\label{lem:kTreePreservedUnderMergeTwo}
Assume that $B$ is in $\ktree$ and nodes $v_1$ and $v_2$ are at
distance at most two in $S$.
The graph $B'$ obtained from $S$ by merging $v_1$ and $v_2$ in $B$ is also in $\ktree$.
\begin{Sketch}
By definition of $\ktree$, there exists a set of edges $D \subseteq E$ such that $B \setminus D$, denoted by $T$, is acyclic and $|D| \leq k$.
We show how to transform $D$ into $D' \subseteq E'$ such that $B' \setminus D'$, denoted by $T'$, is acyclic and $|D'| \leq k$.
We consider three cases, depicted in \figref{mergeTrees}.

\begin{itemize}
\item
If $e_1, e_2 \notin D$ then $D' = \{ m(e) | e \in D \}$.
\item
Assume that $e_1 \notin D$ and $e_2 \in D$.
If $v_2$ is not reachable from $v_1$ in $T$,
then $D' = \{ m(e) | e \in D \}$, thus $|D'| \leq k$.

If $v_2$ is reachable from $v_1$ in $D$,
there is at most one path from $v_1$ to $v_2$ in $T$,
because $T$ is acyclic.
If the path contains $e_1$, we define $D'$ as before: $D' = \{ m(e) | e \in D \}$.

If the path from $v_1$ to $v_2$ does not contain $e_1$,
let $e_3$ be the first edge on the path from $v_1$ to $v_2$ (see the second case in \figref{mergeTrees}).\footnote{Note
that we cannot use the simple $D'$ definition as before,
because merging $v_1$ and $v_2$ in $T$ to obtain $T'$ creates a cycle that does not involve $e_1$.
We observe that, in this case, the subgraph reachable from $v_1$ through $e_1$ in $T$
remains acyclic after the merge operation, because it is disjoint from the subtree of $v_2$.
Thus, $e_1$ need not be removed from $T$.}
To obtain $D'$ from $D$, we remove $e_2$ and add $e_3$:
$D' = (\{ m(e) | e \in D \} \setminus \{ m(e_2) \})  \cup \{ m(e_3) \}$.
This definition does not increase the size of $D$, because $e_2 \in D$.
\item
Assume that $e_1, e_2 \in D$.
If $v_2$ is not reachable from $v_1$, we can use the simple construction $D' = \{ m(e) | e \in D \}$.
It follows that $|D'| = |D| - 1$, because both $e_1$ and $e_2$ are mapped to the same edge $e'$.

If $v_2$ is reachable from $v_1$, let $e_3$ be the first edge on the path.
We define $D' = \{ m(e) | e \in D \} \cup \{ m(e_3) \}$ (see the third case in \figref{mergeTrees}).
Same construction applies when $v_1$ or $v_2$ are reachable from $v_0$.
\end{itemize}

\begin{figure*}
\begin{center}
\begin{tabular}{|c|c|c|}
\hline
& $T$ & $T'$ \\
\hline
$e_1, e_2 \notin D$
&
\includegraphics[width=2cm,angle=-90]{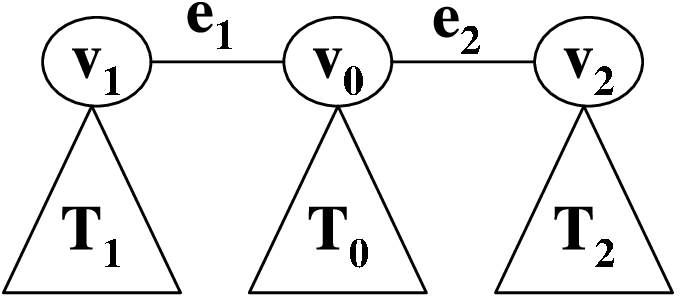}
&
\includegraphics[width=2cm,angle=-90]{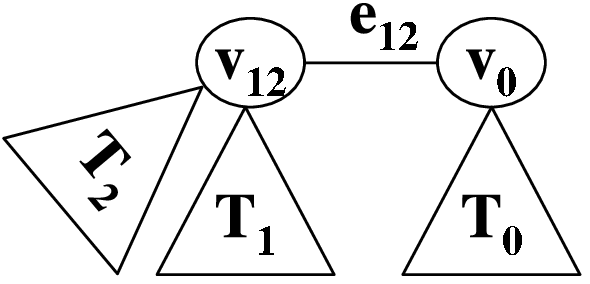}
\\
\hline
$e_1 \notin D, e_2 \in D$
&
\includegraphics[height=4cm,angle=-90]{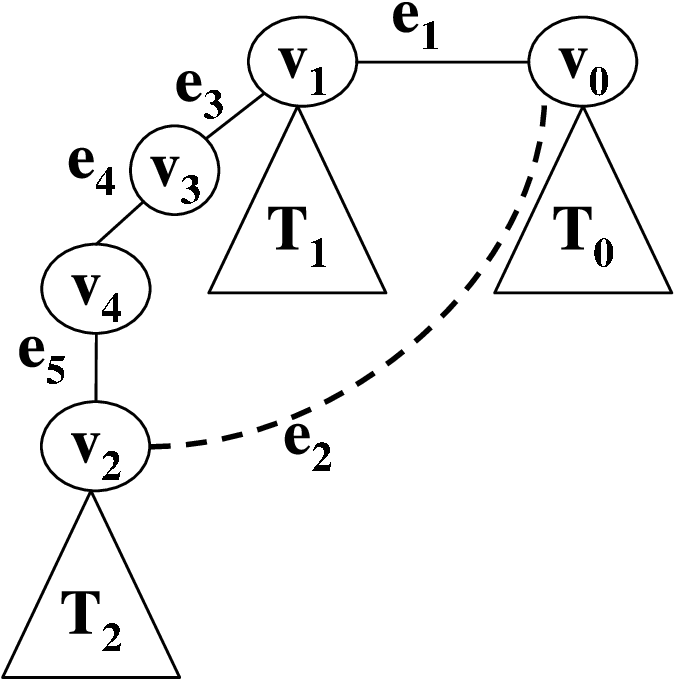}
&
\includegraphics[width=2.5cm,angle=-90]{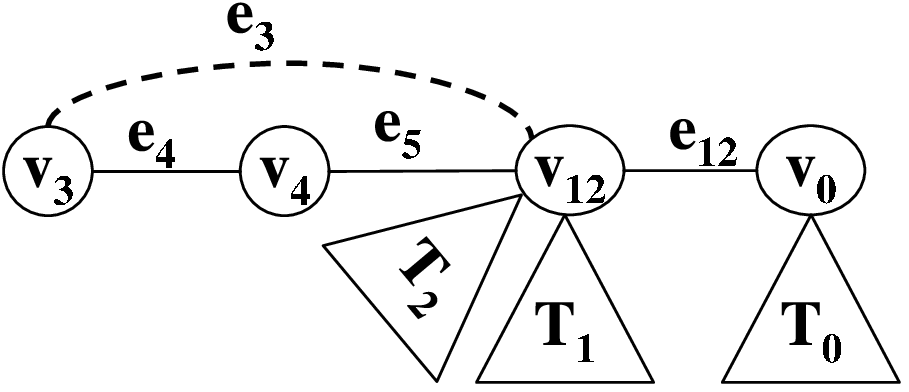}
\\
\hline
$e_1, e_2 \in D$
&
\includegraphics[height=4cm,angle=-90]{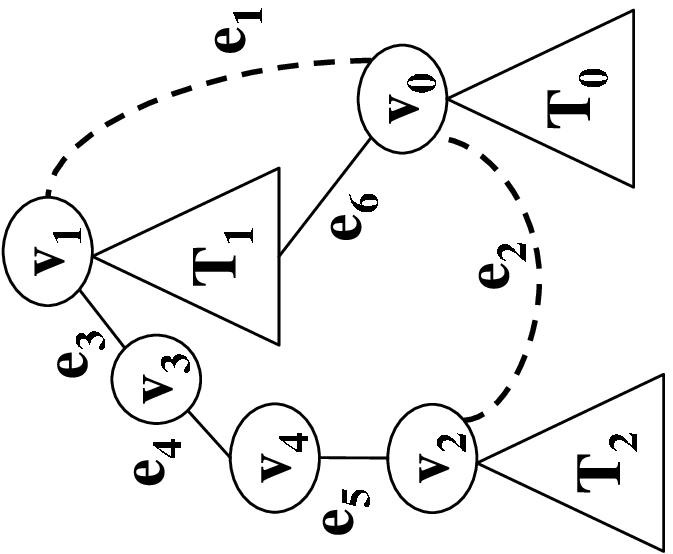}
&
\includegraphics[height=5cm,angle=-90]{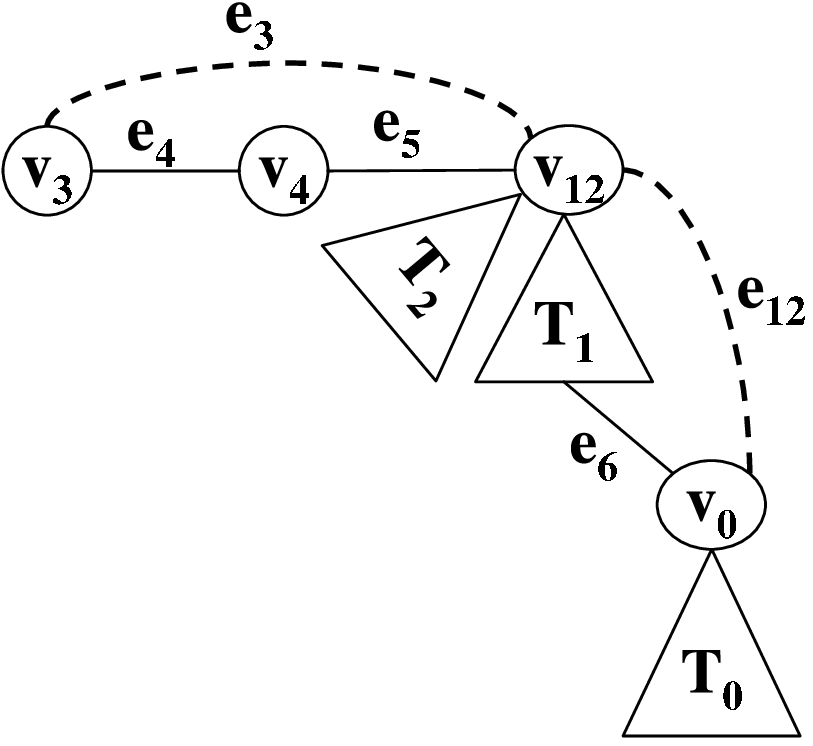}
\\
\hline
\end{tabular}
\end{center}
\caption{\label{Fig:mergeTrees} Merge operation on $\ktree$-graphs. Dotted
lines represent additional edges, i.e., edges of a $\ktree$-graph that do not belong to the tree.
The node $v_{12}$  and the edge $e_{12}$ in $T'$ result from merging the nodes $v_1$ and $v_2$, and the edges $e_1$ and $e_2$ in $T$.}
\end{figure*}
\end{Sketch}
\end{Lemma}

\begin{Lemma}\label{lem:mergeAyah}
Assume that $S$ is  in  $\Ayah{k}$  and nodes $v_1$ and $v_2$ are
at distance at most $2$ in $S$. The graph $S'$ is obtained from $S$ by
merging $v_1$ and $v_2$ in $S$.
We show that $S'$ is in $\Ayah{k}$.
\begin{Sketch}
To show that $S' \in \Ayah{k}$, it is sufficient to show that $\gaifman{S'} \in \ktree$.
Note that merging the nodes of $\gaifman{S}$ that correspond to $v_1$ and $v_2$ in $\gaifman{S_j}$,
results in $\gaifman{S'}$.
The restriction on the disequality patterns in $\LRPtwo$
guarantees that the nodes $v_1$ and $v_2$ are at distance at most $2$ in $\gaifman{S}$.
$\gaifman{S} \in \ktree$, because $S \in \Ayah{k}$, and
using \lemref{kTreePreservedUnderMergeTwo}, we get that $\gaifman{S'} \in \ktree$.
\end{Sketch}
\end{Lemma}

\begin{Lemma}\label{lem:additionAyah}
Assume that $S$ is  in  $\Ayah{k}$  and nodes $v_1$ and $v_2$ are
at distance at most $1$ in $S$. The graph $S'$ is obtained from $S$ by adding an edge
from $v_1$ to $v_2$.
We show that $S'$ is in $\Ayah{k}$.
\begin{Sketch}
Distance at most $1$ between $v_1$ and $v_2$ means that there is an edge.
Addition of edges to $S$ in parallel to existing edges does not affect the $\gaifman{S}$,
and self-loops are ignored in the $\ktree$-graph.
\end{Sketch}
\end{Lemma}

\subsection{Construction}

\begin{SThe}{the:AyahModel}
Let $\phi \in \NLBPtwo$ as defined above.
If $\phi$ is satisfiable, then $\phi$ is satisfiable by a graph in $\Ayah{k}$,
where $k = 2 \times n \times |C| \times m$,
$m$ is the number of constraints in $\phi_{\Diamond}$,
$|C|$ is the number of constants in the vocabulary, and
for every regular expression that appears in $\phi_{\Diamond}$ there
is an equivalent automaton with at most $n$ states.
\begin{Proof}
Let $S$ be a model of $\phi$ : $S \models \phi$.
We construct a graph $S'$ from $S$ and show that $S' \models \phi$ (\lemref{resModel}) and
$S' \in \Ayah{k}$ (\lemref{resAyah}).
The construction is defined as follows:
\begin{enumerate}
\item
For each constraint $i$ in $\phi_{\diamond}$,
    identify the shortest witness $W_i$ in $S$,
    and split it into a witness $W'_i$.
\item The graph $S_0$ is a union of all $W'_i$'s, in which the
    nodes labelled with the (syntactically) same constants are merged.
\item
    Apply all enabled merge operations and all enabled edge-addition operations in any order,
    producing a sequence of distinct graphs $S_0, S_1, \ldots, S_m$,
    until $S_m$ has no enabled operations.
\item The result $S' = S_m$.
\end{enumerate}
The process described above terminates after a finite number of steps:
because there is a finite number of nodes in $S_0$,
each merge operation reduces the number of nodes,
and only finitely many edges can be added to a graph of a given size.
\end{Proof}
\end{SThe}

\begin{Lemma}\label{lem:resModel}
$S' \models \phi$
\begin{Sketch}
The graphs generated by the process above are related to each other by
different homomorphism relations (\defref{homomorphism}),
as depicted in \figref{trans}.

First, each step of the process can be seen as a transformation $t_j$ from $S_{j-1}$ to $S_j$,
which is defined by an operation applied at step $j$.
That is, $t_j$ is either a merge operation or an edge-addition operation.
It is easy to see that both operations are homomorphisms,
and therefore, each $t_j$ is a homomorphism, for all $j$.

Second, we define a mapping $f_j$ from $S_0$ to $S_j$ as a composition $t_j \circ \ldots t_0$;
the mapping $f_j$ is a homomorphism, because it is a composition of homomorphisms.
By construction, $S_0$ contains a witness for each reachability constraint in $\phi_{\Diamond}$,
thus,  $S_0 \models \phi_{\Diamond}$.
By \lemref{homoPres} and the existence of a homomorphism $f_j$ from $S_0$ to $S_j$
we get that $\phi_{\Diamond}$ is satisfied by $S_j$, because it is satisfied by $S_0$.
In particular, $S' \models \phi_{\Diamond}$.

Third, from the definition of witness splitting and \lemref{mergeHomo},
we can show that there is a homomorphism $h_0$ from $S_0$ to $S$.

Finally, we define a mapping $h_j$ from $S_j$ to $S$, for all $j = 1, \ldots, m$;
$h_j$ is a homomorphism  (by \lemref{mergeHomo} and induction on $j$).
By \lemref{homoPres} and the existence of a homomorphism $h_j$ from $S_j$ to $S$
we get that $\phi_{\Box}$ is satisfied by $S_j$, because it is satisfied by $S$.
In particular, $S' \models \phi_{\Box}$.

The graph $S'$ is obtained upon termination, and therefore it satisfies $\phi_{=}$ and $\phi_{\rightarrow}$.
This completes the proof that $S' \models \phi$.
\begin{figure}
\begin{center}
\includegraphics[width=4cm,angle=-90]{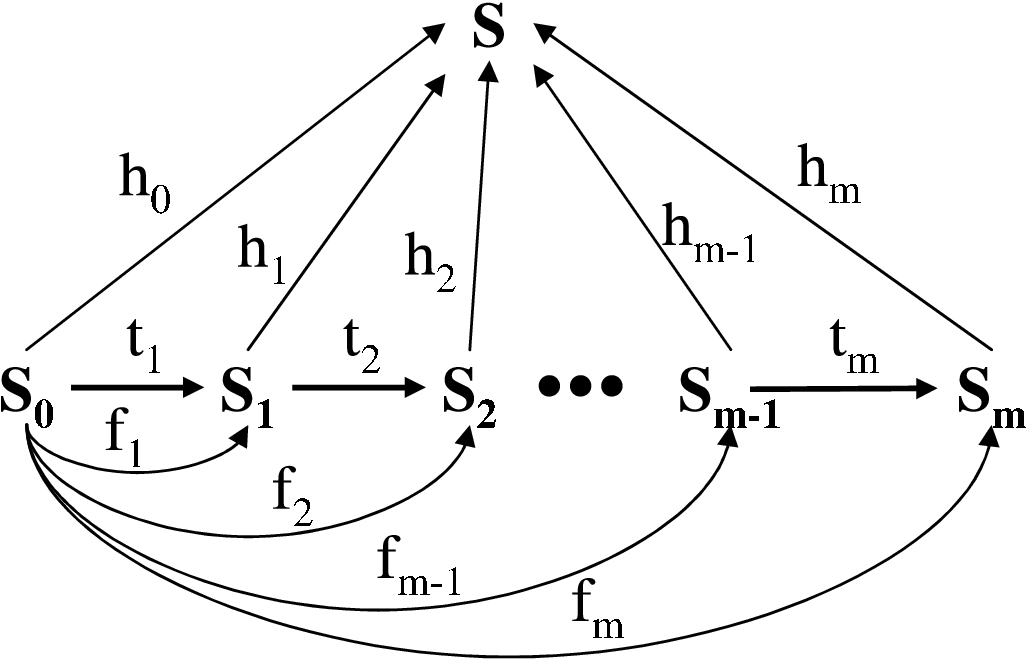}
\end{center}
\caption{\label{Fig:trans} Construction and homomorphisms in the proof of decidability.}
\end{figure}

\end{Sketch}
\end{Lemma}

\begin{Lemma}\label{lem:resAyah}
For all j, $S_j \in \Ayah{k}$, where $k = 2 \times n \times |C| \times m$,
$m$ is the number of constraints in $\phi_{\Diamond}$,
$|C|$ is the number of constants in the vocabulary, and
for every regular expression that appears in $\phi_{\Diamond}$ there
is an equivalent automaton with at most $n$ states.
\begin{Sketch}
If a node is visited more than once in the same state of the automaton, the path can be
shortened by removing the part traversed between the two visits.
Thus, the shortest witness visits a node labelled with a constant at most $n$ times.
In the worst case, each visit enters and exists the node with a different edge,
producing $2 \times n$ edges adjacent to a node labelled with a constant.

Recall that $S_0$ consists of $m$ short witnesses.
The witness paths are disjoint, except maybe the nodes labelled with constants.
Thus, every constant has at most $2 \times n \times m$ adjacent edges.

By construction, every cycle must go through a constant.
To break all cycles in $S_0$ (and its Gaifman graph), it is sufficient to remove all the edges
adjacent to nodes labelled with constants,
i.e., $k = 2 \times n \times m \times |C|$ edges.
It follows that $S_0 \in \Ayah{k}$ (the bound is not tight).

By inductive hypothesis, $S_{j-1} \in \Ayah{k}$.
If $t_j$ is a merge operation, then $S_j \in \Ayah{k}$ by \lemref{mergeAyah}.
If $t_j$ is an edge-addition operation, then $S_j \in \Ayah{k}$ by \lemref{additionAyah}.
\end{Sketch}
\end{Lemma}

\subsection{Translation from $\LRP$ to MSO}\label{app:LRP2MSO}

Every regular expression $R$ can be translated into an MSO formula $\phi_R(x,y)$,
that describes the paths from $x$ to $y$ labelled with $w$, for every word $w$ in $R$.
To encode the Kleene star expression, we use a least fixpoint operation, expressible in MSO.

\begin{Lemma}\label{lem:regularMSO}
Every routing expression $R$ can be translated into an MSO formula $\phi_R(x,y)$
with two (first-order) free variables $x$ and $y$ such that
for every graph $S$ and nodes $a,b \in S$, there is an $R$-path from $a$ to $b$ if and only if $S, a, b \models \phi_R(x,y)$.
\begin{Sketch}
For atomic regular expressions and concatenation, we define $\phi_R(x,y)$ as follows:
\[
\begin{array}{ll}
\phi_R(x,y) & \eqdef
\left \{ \begin{array}{ll}
                         f(x,y)   &    \mbox{if}~R~\mbox{is}~\Edge{f} \\
                         f(y,x)   &    \mbox{if}~R~\mbox{is}~\BEdge{f} \\
                         \neg(c = x) \land (x = y)                      &    \mbox{if}~R~\mbox{is}~\neg c\\
                         u(x) \land (x = y)                            &    \mbox{if}~R~\mbox{is}~u\\
                         \neg u(x) \land (x = y)                      &    \mbox{if}~R~\mbox{is}~\neg u\\

        \end{array} \right .
        \\
\phi_{R_1.R_2}(x,y) & \eqdef  \exists z . \phi_{R_1}(x,z) \land \phi_{R_2}(z,y)\\
\end{array}
\]
The formula $\phi_{R^*}(x,y)$ holds when the minimal set $Y$ that contains
$x$ and closed under $R$, contains $y$. Formally, we define
\[
\begin{array}{ll}
\phi_{R^*}(x,y) \eqdef &
\exists Y . (y \in Y) \land Q(x, Y) \land \forall Y' . Q(x, Y') \implies Y \subseteq Y'
\end{array}
\]
where $Q(x, Z)$ is
$(x \in Z) \land \forall x', y' . (x' \in Z) \land \phi_R(x',y') \implies (y' \in Z)$.
\end{Sketch}

\end{Lemma}

Using the translation above, it is easy to translate a general $\LRP$ formula to an \emph{equivalent} MSO formula.
For a formula  $\phi \in \LRP$, we define the translation $\Tr$ to MSO formulas inductively:
\[
\begin{array}{ll}
\Tr(\negFormula{c}{R}{p}) & \eqdef \forall x . \phi_R(c, x) \implies p(x)\\
\Tr(\phi_1 \land \phi_2) & \eqdef \Tr(\phi_1) \land \Tr(\phi_2)\\
\Tr(\neg \phi_1 ) & \eqdef \neg \Tr(\phi_1)
\end{array}
\]
\begin{Lemma}\label{lem:TrMSO}
For all $\phi \in \LRP$ and all graphs $S$, $S \models \phi$ iff $S \models \Tr(\phi)$.
\end{Lemma}

\subsection{Reduction from MSO on Ayah Graphs to MSO on Trees}

The satisfiability problem of MSO logic on Ayah graphs can be reduced to
the satisfiability problem of MSO logic on trees, which is decidable, using a classical result due to Rabin~\cite{Rabin}.
This reduction completes the proof of decidability of $\LRPtwo$.
Furthermore, it provides a constructive way to check satisfiability $\LRPtwo$ formulas, using an existing
decision procedure for MSO on trees, MONA~\cite{KlaEtAl:Mona}.

The reduction consists of two satisfiability-preserving translations:
(i)~translation $\Tr_1$ from MSO on Ayah graphs
to MSO on directed forests, and (ii)~translation $\Tr_2$ from MSO on directed forests to MSO on
infinite binary trees.
\begin{Lemma}\label{lem:TrMona}
There are translations $\Tr_1$ and $\Tr_2$ between MSO-formulas such that
for every MSO-formula $\phi$,
there exists a graph $S \in \Ayah{k}$ that satisfies $\phi$
if and only if there exists a binary tree $S'$ such that $S' \models (\Tr_1 \circ Tr_2) (\phi) $.
\end{Lemma}
In this paper, we  describe only the first translation $\Tr_1$, and omit the second (standard) translation, $Tr_2$.

\paragraph{Direct Forests}
A directed forest is a set of directed trees,
defined over a single binary relation symbol $E$,
and arbitrary unary relation symbols and constant symbols.
An $E$-edge from node $u_1$ to node $u_2$ means that $u_2$ is a child of $u_1$ in the tree.
Each node in $S$ has at most one incoming $E$ edge.
Let $T$ denote a set of all directed forests.

For every MSO formula $\phi$ over the vocabulary $\voc$,
we construct an MSO formula $\Tr_1(\phi)$ over the vocabulary $\voc'$
such that $\phi$ is satisfiable on $\Ayah{k}$ if and only $\Tr_1(\phi)$ is satisfiable on $F$.

The new vocabulary $\voc'$ contains $E$, all unary relation symbols and constants of $\voc$,
and new unary relation symbols and constants:
\[
\begin{array}{l}
\voc' = \{E\} \cup U \cup C \cup  \{ F_f, B_f \mid f \in F\} \cup \\
\{ c^f_i, d^f_i, b^f_i \mid i = 1, \ldots, k~\mbox{and}~f \in F\}
\cup \{ L^f \mid f \in F\}
\end{array}
\]

The translation $Tr_1$ is defined inductively in the graph of $\phi$,
where the only interesting part is the translation of a binary relation formula:
\[
\begin{array}{l}
\Tr_1( f(v_1, v_2) ) = (E(v_1,v_2) \land F_f(v_2)) \lor (E(v_2, v_1) \land B_f(v_1)) \lor \\
\Lor_{i=1, \ldots, k} (( c_i^f = v_1 \land d_i^f = v_2 ) \lor ( c_i^f = v_2 \land d_i^f = v_1 \land b_i^f = v_2 ))\\
\lor ( v_1 = v_2 \land L^f(v_1) )
\end{array}
\]

We describe the intuition behind this translation.
Given a graph $S$ in $\Ayah{k}$, we can remove all self loops and at most $k$ additional edges
from the Gaifman graph of $S$ to obtain an acyclic (undirected) graph, denoted by $U$.
It is easy to transform the undirected graph $U$ into a directed forest $D$.
The idea of the translation $\Tr_1$ is to encode edge labels
and edge directions of $S$ using new unary relation symbols and constants in $D$.

For each binary relation symbol $f \in F$,
we introduce two unary relation symbols $F_f$ and $B_f$, denoting forward and backward $f$-edge.
There is an $f$-edge from $u$ to $u'$ in $S$ if there is a (forward) $E$-edge from $u$ to $u'$ in $D$,
and $u'$ is labelled with $F_f$.
Similarly, there is an $f$-edge from $u$ to $u'$ in $S$ if there is a (backward) $E$-edge from $u'$ to $u$ in $D$,
and $u$ is labelled with $B_f$.

In addition, we record the self-loops that appear in $\gaifman{S}$, but not in $D$.
Each such self-loop can correspond to several self-looping edges in $S$, labelled by different labels.
We record a self-loop $f$ on a node $u$ in $S$ by labelling the node $u$ in $D$ with a unary relation symbol $L^f$.

Also, we encode other edges that appear in $\gaifman{S}$ but not $D$ (at most $k$ of those).
Each such edge can correspond to several parallel edges in $S$, possibly in different directions,
labelled with different labels.
For each such edge $f$, we label the source and the target nodes with $c_i^f$ and $d_i^f$,
respectively. If an $f$-edge is bi-directional, we label the target with an additional constant symbol $b_i^f$.

\begin{Lemma}
Let $\phi$ be an MSO formula.
There is a graph $S \in \Ayah{k}$ such that $S \models \phi$ if and only if there is a graph $S' \in T$
such that $S' \models \Tr(\phi)$.
\end{Lemma}

}
}
\end{document}